\documentclass[12pt,twoside]{article}
\usepackage{correl}

\def\TheTitle{Dynamical Mean-Field Theory of Disordered Electrons: Coherent Potential Approximation and Beyond}
\def\TheHeading{DMFT \& CPA}
\def\TheAuthor{V\'aclav Jani\v s}
\def\TheInstitute{Institute of Physics,  Czech Academy of Sciences,  Na Slovance 2, CZ-182 21 Praha, Czech Republic}
\def\TheAddress{}
\def\TheChapter{5}                       
 \def\veck{\mathbf k}
\def\vecq{\mathbf q}

\newcommand{\be}{\begin{equation}}
\newcommand{\ee}{\end{equation}}
\newcommand{\bml}{\begin{multline}}
\newcommand{\eml}{\end{multline}}
\newcommand{\bal}{\begin{align}}
\newcommand{\eal}{\begin{align}}

\begin{document}
\MakeTitle           
\section{Introduction}

Impurities are ubiquitous in real materials and are non-negligible in reliable realistic calculations of the low-temperature properties of solids. The impurities are randomly distributed on a macroscopic scale and their impact on thermodynamic, spectral and transport properties of solids is an important topic of experimental and theoretical research.  Advanced experimental techniques allow now for a rather precise determination of the chemical composition of heterogenous materials, which on the other hand, increases demands on the precision of the theoretical description of materials with randomness.  

Lax \cite{Lax:1951aa,Lax:1952aa} was the first who simulated scatterings on a random potential by a self-consistently determined homogeneous effective medium. The idea of an effective medium went far beyond a rigid band or a virtual crystal approximation standardly used at that time. The coherent imbedding into a homogeneous environment, when transferred to the context of random alloys,  has become the corner-stone of what later has become known as the coherent potential approximation (CPA). The idea of Lax was further extended by Davies and Langer by multiple single-site scatterings  \cite{Davies:1963aa}.  

The equations for the genuine coherent potential approximation were introduced independently by Soven  \cite{Soven:1967aa} and Taylor \cite{Taylor:1967aa}.  The multiple-scattering approach was applied in Ref.~\cite{Soven:1967aa} to electrons on random lattices while in Ref.~\cite{Taylor:1967aa} to lattice vibrations  of imperfect crystals. The method of coherent potential was extensively studied and applied in various situations during the late sixties and in the seventies of the last century.   The progress in the description of random media via multiple single-site scatterings was made possible by two principal theoretical developments. First, sophisticated many-body perturbation techniques using Green functions made it possible to avoid a cumbersome description of random media via an inhomogeneous  differential Schr\"odinger equation. Second, the development of computers capable of determining numerically exactly reasonably large clusters opened the way to the application of the coherent potential methodology to real materials beyond the model level.       

The coherent potential approximation has several attractive features. It was shown to be the best single-site approximation, \cite{Velicky:1968aa} and the coherent potential has the proper analytic properties in the complex energy plane consistent with causality of the averaged Green function \cite{Mueller-Hartmann:1973aa,Ducastelle:1973aa}. Finally, transport properties of disordered systems can also be determined within this single-site approximation \cite{Velicky:1969aa}. The early approaches to CPA were reviewed in Ref.~\cite{Elliott:1974aa}. 

The coherent potential approximation remained for long singled out from other approximations due to its analytic structure and accuracy in the determination of thermodynamic and transport properties. Direct cluster extensions of the single-site multiple-scatterings failed in keeping  causality of the Green function \cite{Capek:1971aa}. At that time the only causal cluster extension, the so-called traveling cluster approximation \cite{Mills:1973aa,Mills:1978aa}, was unhandy for applications in realistic settings. 

A new impetus in understanding of CPA in a broader context came in the late eighties and early nineties of the last century with the concept of the Dynamical Mean-Field Theory (DMFT).  First, a functional-integral generalization of CPA  enabled to understand the concept of coherent potential as a single-site approximation with self-consistently summed single-loop contributions of the many-body perturbation expansion of the thermodynamic potential \cite{Janis:1986aa,Janis:1989aa}. Second, these diagrams were then showed to determine the limit of the exact solution of models of interacting electrons to the infinite-dimensional hypercubic lattice \cite{Metzner:1989aa}. Consequently, CPA then appeared to be an exact solution of models of the disordered Fermi lattice gas in the limit to infinite dimensions within DMFT \cite{Janis:1989aa,Vlaming:1992aa}.  Since then, CPA is understood as a special case of DMFT applied to disordered systems.  Not only this, DMFT and CPA are interconnected in systematic ways to improve upon these local approximations built on many-body diagrammatic approaches.

\section{Quantum mechanics of a particle in a static random environment}

\subsection{Lattice model with a random atomic potential}

Both electron correlations and randomness in configurations of impurities or in the chemical composition are always present to some extent in real materials. It is wise to separate them first to understand their individual  impact on the behavior of the electrons.  Hence, the easiest model of disorder in metals, crystalline solids with available conduction electrons, is a Fermi gas of moving light particles scattered on heavy, immobile ions, the atomic potential of which fluctuates from site to site. Since there are no electron correlations present, we have a quantum mechanical problem of a test particle scattered on randomly distributed atomic potential of ions ordered in a regular lattice structure.  The generic quantum mechanical Hamiltonian of such an electron can be written as 
\begin{equation}\label{eq:H-QM}
\widehat{H} = \sum_{nm}\left\lvert m\right\rangle W_{mn}\left\langle n\right\rvert + \sum_{n} \left\lvert n\right\rangle V_{n}\left\langle n\right\rvert = \widehat{W} + \widehat{V}\, ,
\end{equation}
where  $W_{mn} = W(\vec{R}_{m} - \vec{R}_{n})$ with $W_{nn} = 0$ is the hopping amplitude of the electron between lattice sites $\vec{R}_{m}$ and $\vec{R}_{n}$ and $\lvert n \rangle, \lvert m \rangle$ stand for Wannier states at the respective lattice sites.  One usually resorts to the hopping between only the nearest neighbors. Local potential $V_{n}$ acquires values due to the atomic occupation of the lattice site $\vec{R}_{n}$.  In case of a binary alloy with atoms of type $A$ and $B$ the probability distribution of the atomic potential is 
\be\label{eq:random-dist}
g(V) = x_{A}\delta(V - V_{A}) + x_{B}\delta(V - V_{B}) \,,
\ee
with  $x_{A} =N_{A}/N = c$ and $x_{B} = 1 -  c$ are densities of atoms $A$ and $B$, respectively, and $N$ being the number of the electrons/lattice sites. We assume that the lattice sites are occupied independently according to distribution $g(V)$ from Eq.~\eqref{eq:random-dist}.

Fluctuations in the values of the atomic potential strongly influence the motion of the electron. Since the operators of the hopping $\widehat{W}$ and the potential $\widehat{V}$ do not commute, the Schr\"odinger equation for the electron in a randomly distributed scattering potential is not exactly solvable for extended systems. Eigenvalues of the Hamiltonian from Eq.~\eqref{eq:H-QM} are random numbers. We hence have to determine the distribution of the eigenvalues of the random Hamiltonian so that to make conclusions about the behavior of the test particle in the random environment.

The fundamental quantity in the description of the quantum particle is  the resolvent defined for an arbitrary complex energy $z$ outside the real axis    
\be
G_{mn}(z) = \left\langle m\left\lvert \left[z \hat{1} -  \widehat{W} - \widehat{V} \right]^{-1}\right\rvert n\right\rangle \,.
\ee
The distribution of the eigenenergies (density of states) then is 
\be\label{eq:DOS-def}
\rho(E) = - \frac 1{\pi V} \sum_{n}\Im G_{nn}(E + i0^{+}) \,,
\ee
with $V= Nv$, and $v$ is the volume of the elementary cell. This distribution generally depends on the size of the random system as well as on the boundary conditions for solving the Schr\"odinger equation. This dependence is removed by configurational averaging.

\subsection{Configurational averaging: coherent potential and T-matrix operator}

Configurational averaging \index{configurational averaging} is a tool for reinstalling translational invariance in random systems. It enables us to develop systematic approximations to the physical quantities of interest. Randomness introduces fluctuations into the physical quantities, since the eigenvalues of the random Hamiltonian are spread on an interval of the real axis. Summing over configurations takes account of the fluctuations only on average and hence not all averaged quantities are relevant. For instance, the averaged Hamiltonian, the energy, is only of a little value. Moreover, products of random variables differ from the product of their averages and vertex corrections for products of random variables must be introduced. 

Each averaged quantity is then characterized by a translationally invariant function containing on average the impact of randomness on this function. It is the coherent potential for the averaged resolvent defined from the following equation   
\be\label{eq:CP-def}
 \left\langle  \mathbb G \right\rangle_{av} =  \widehat{G} = \left[z\widehat{1} - \widehat{\sigma}(z) - \widehat{W}  \right]^{-1}\,.
\ee
The exact coherent potential \index{coherent potential} is generally a non-local operator on the lattice  $\widehat{\sigma}(z) = \sum_{n,m} \left\lvert n\right\rangle \sigma_{nm}(z)\left\langle m\right\rvert $, but we resort only to single-site approximations with a diagonal coherent potential $\widehat{\sigma}(z) = \sum_{n} \left\lvert n\right\rangle \sigma_{n}(z)\left\langle n\right\rvert $.

The coherent potential contains the fluctuations due to the random character of the scattering potential only in an averaged manner. We introduce a configurationally-dependent  T-matrix operator $\mathbb T (z)$ containing the fluctuations missing in the coherent potential
\be
\mathbb G(z) = \left\langle \widehat{G}(z)\right\rangle_{av} +  \left\langle \widehat{G}(z)\right\rangle_{av} \mathbb T (z) \left\langle \widehat{G}(z)\right\rangle_{av} \,. 
\ee
The T matrix is  generally a nonlocal operator and, similarly to the coherent potential, we can introduce local T matrices 
\be
\mathbb T_{n}(z) = \frac{V_{n} - \sigma_{n}(z)}{1 - \left(V_{n} - \sigma_{n}(z)\right)G_{nn}(z)}
\ee
that depend on the lattice coordinate $\vec{R}_{n}$ only and contain all multiple single-site scatterings on the fluctuations of the atomic potential $V_{n}$ relatively with respect to the coherent potential of the effective medium $\sigma_{n}(z)$.

The full  T matrix \index{T matrix} can be represented via the local ones and wave operators $\mathbb Q_{n}(z)$ as 
\begin{align}
\mathbb T(z) &= \sum_{n}\mathbb T_{n}(z)\left[\widehat{1} + \left\langle \mathbb G(z)\right\rangle_{av}\sum_{m\neq n}\mathbb Q_{m}(z) \right] \,,
\\
 \mathbb Q_{n}(z) &= \mathbb T_{n}(z) + \left[\widehat{1} + \left\langle \mathbb G(z)\right\rangle_{av}\sum_{m\neq n}\mathbb Q_{m}(z) \right]  \,.
 \end{align}
 Successive exclusion of the wave operators leads to a representation of the T matrix via  a multiple-scattering series with the local T matrices connected by the  averaged resolvent
\begin{multline}\label{eq:TMatrix-expansion}
\mathbb T(z) = \sum_{n}\mathbb T_{n}(z) + \sum_{n\neq m}\mathbb T_{n}(z)\left\langle \mathbb G_{nm}(z)\right\rangle_{av}\mathbb T_{m}(z) 
\\
+ \sum_{n\neq m\neq l}\mathbb T_{n}(z)\left\langle \mathbb G_{nm}(z)\right\rangle_{av}\mathbb T_{m}(z)\left\langle \mathbb G_{ml}(z)\right\rangle_{av} \mathbb T_{l}(z) +\ldots \quad\,.
\end{multline}
This is a typical excluded-volume problem that is difficult to solve beyond first few terms of the series. 

The averaged T matrix vanishes in the exact solution. It means, that the  coherent potential captures all the fluctuations of the random atomic potential. Since we resorted to a single-site coherent potential, we cannot guarantee vanishing of the full T matrix but only of its local part. Then vanishing of the local T matrix   
\be\label{eq:T-matrix-Soven}
\left\langle \mathbb T_{n}(z) \right\rangle_{av } =  \left\langle  \frac{V_{n} - \sigma(z)}{1 - (V_{n} - \sigma(z))\left\langle G_{nn}(z)\right\rangle_{av}}\right\rangle_{av} =  0 
\ee
is the \index{Soven equation} Soven equation  for the coherent potential $\sigma_{n}(z)=\sigma(z)$.  Averaging restores translational invariance, hence the coherent potential is independent of the lattice coordinate. 

Soven equation for the coherent potential can be solved only iteratively. Due to the correct analytic properties of the coherent potential we can guarantee convergence of the following iteration procedure for real energies  $\lim_{l\to\infty}\sigma^{(l)}(E_{+}) = \sigma(E_{+})$
\be
\Im\sigma^{(l + 1)}(E_{+}) = \left[1 - \frac{\lvert G^{(l)}(E_{+})\rvert^{2}}{\left\langle \lvert G^{(l)}(E_{+})\rvert^{2}\right\rangle}\right]\left\langle \frac 1{\lvert 1 + G^{(l)}(E_{+})\left( \sigma^{(l)}(E_{+}) - V_{n}\right)\rvert^{2}}\right\rangle_{av} \Im\sigma^{(l)}(E_{+})\,, 
\ee
where we denoted  $E_{+}= E + i0^{+}$, $\left\langle \lvert G(z)\rvert^{2}\right\rangle = \Im G(z)/\Im \sigma(z) = \int d\epsilon \rho(\epsilon) \lvert z - \epsilon - \sigma(z)\rvert^{-2}$ and $\rho(\epsilon)$ is the density of states of the electrons on the homogeneous lattice. Negative sign of the imaginary part of the coherent potential is guaranteed during the iterations. The analytic properties of CPA must not be broken during the iterative process of numerical solutions in order to stay within the physical phase space.

\section{Many-body approach to disordered electron systems}

\subsection{Thermodynamic limit and translational invariance}

The concept of an \index{effective medium} effective medium and a coherent potential was derived with quantum mechanics of particles, that is, for Fermi or Bose gases without inter-particle interactions. The construction of the best local approximation is an appealing approach that can be used in a broader context, namely in the statistical mechanics of many-body systems.    

The equilibrium properties of macroscopic many-body systems are extracted from the thermodynamic limit. It means that the volume $V$ of the system is sent to infinity. The differences between positions of the individual sites vanish and translational invariance is restored in the \index{thermodynamic limit} thermodynamic limit. We can use the Fourier transform from the direct lattice to momentum space and use the Bloch waves as the elementary quantum states that form a complete orthonormal basis of the state states with which we can describe the random system in the thermodynamic limit.  

We use second quantization and extend the quantum-mechanical Hamiltonian, Eq.~\eqref{eq:H-QM}, to the \index{Anderson disordered model} Anderson disordered model in the Fock space by means of creation and annihilation fermionic operators $c_{i}^{\dagger }$, $c^{\phantom{\dagger}}_{i}$ respectively
\begin{equation}\label{eq:Anderson-Hamiltonian}
  \widehat{H} = - t \sum_{<ij>}c_{i}^{\dagger }c^{\phantom{\dagger}}_{j }+\sum_i V_i c^{\dagger
}c^{\phantom{\dagger}}_{i} 
 =\sum_{\mathbf{k}}
   \epsilon(\mathbf{k})c^{\dagger}(\veck) c(\veck)  
  + \sum_i  V_i c^{\dagger}_{i} c^{\phantom{\dagger}}_{i}\,,
\end{equation}
where $\epsilon(\veck) = \sum_{i}W_{i0}\exp\{i\vec{R}_{i}\cdot\veck\}$ is the dispersion relation of the Fermi gas on the lattice, $c^{\dagger}(\veck)= V^{-1}\sum_{i}c^{\dagger}_{i}\exp\{i\vec{R}_{i}\cdot\veck\}$, and $\langle ij\rangle$ denotes nearest-neighbor lattice sites with coordinates $\vec{R}_{i}$ and $\vec{R}_{j}$.  

The existence of the equilibrium state in the thermodynamic limit is conditioned by validity of the \index{ergodic hypothesis} ergodic hypothesis, it means, that the particle passes almost everywhere in the phase space after sufficiently long time. Then the spatial averaging equals configurational averaging, at least for local quantities that can be proved to possess the so-called \index{self-averaging property} self-averaging property. For example, 
\begin{equation}\label{eq:}
\rho(E) = - \frac 1{\pi V}\sum_{i}\Im G_{ii}(E_{+}) = -\frac 1\pi\left\langle \Im G_{ii}(E_{+})\right\rangle_{av} = -\frac 1{\pi V}\sum_{\veck} \Im G(\veck,E_{+})
\end{equation}
holds in the thermodynamic limit. The averaged Green function in the thermodynamic limit then can be represented as
\be\label{eq:Gav-def}
\left\langle\left\langle\mathbf{k}\left| \frac 1{z\widehat{1} -
          \widehat{H}}\right| \mathbf{k}'\right\rangle\right\rangle_{av} = G(\veck,z)\delta(\mathbf{k} - \mathbf{k}') = \frac
  {\delta(\mathbf{k} - \mathbf{k}')}{z - \epsilon(\mathbf{k}) -
    \Sigma(\mathbf{k},z)}\,,
\ee
with $\lvert \veck \rangle = c^{\dagger}(\veck)\lvert \Omega\rangle$ and $\lvert \Omega\rangle$ being the vacuum (cyclic) vector in the Fock space. The delta function in the numerator stands for momentum conservation in translationally invariant systems. The self-energy $\Sigma(\mathbf{k},z)$ contains the entire contribution from the random potential to the one-particle propagator $G(\veck,z)$. It is a many-body generalization of the coherent potential. The thermodynamic limit and the ergodic hypothesis not only restore translational invariance in the random system but also allow us to use the perturbation expansion in the inhomogeneous/random potential so that configurational averaging can be performed  term by term in the perturbation expansion. We can then work only with the averaged Green functions and use  the many-body diagrammatic and renormalization techniques of homogeneous systems.

\subsection{Green functions and relations between them}
\label{sec:GF}
 
The fundamental tool for obtaining quantitative results in disordered systems is a renormalized perturbation theory in the random potential. The perturbation theory works only with translationally invariant averaged Green functions and its basic object is the one-particle Green function from Eq.~\eqref{eq:Gav-def}.  It contains the necessary information about the equilibrium thermodynamic and spectral properties. If we are interested in the response to weak perturbations we need to take into account also the averaged two-particle Green function. If we remove the delta function due to conservation of the total momentum we can define the two-particle Green function with Bloch waves as  
\begin{multline}\label{eq:G2-averaged}
  G^{(2)}_{{\bf k}{\bf
      k}'}(z_1,z_2;\mathbf{q}) =
  \left\langle \left\langle \mathbf{q} + \mathbf{k},\mathbf{k} \left|
        \frac 1{z_1 - \widehat{H}}\otimes \frac 1{z_2 -
          \widehat{H}}\right| \mathbf{k}', \mathbf{q} + \mathbf{k}'
    \right\rangle \right\rangle_{av}  
    \\
    \equiv \left\langle
    \left\langle\mathbf{k} + \vecq \left| \frac 1{z_1 -
          \widehat{H}}\right|\mathbf{k}' + \vecq \right\rangle \left\langle
     \mathbf{k}' \left| \frac 1{z_2 -
          \widehat{H}}\right|  \mathbf{k}\right\rangle
  \right\rangle_{av} \, , 
 \end{multline} 
where $\otimes$ denotes the direct product of operators.  

The full two-particle Green function can further be represented via a vertex function $\Gamma$
\be\label{eq:G2-Gamma}
 G^{(2)}_{{\bf k}{\bf
      k}'}(z_1,z_2;\mathbf{q}) = G_{\mathbf{k} + \vecq}( z_1)G_{
  \mathbf{k}}( z_2)  \\ \times
  \left[\delta(\mathbf{k} - \mathbf{k}') 
 + \ \Gamma_{{\bf k}{\bf k}'}(z_1,z_2;\mathbf{q})
    G_{\mathbf{k}' + \vecq}(z_1) G_{
    \mathbf{k}'}( z_2)\right] \,.
  \ee
The two-particle vertex $\Gamma$ represents a disorder-induced correlation between simultaneously propagated pairs of particles. It measures the net impact of the pair scatterings on the random potential.   

Vertex $\Gamma$ can further be simplified by introducing an irreducible vertex $\Lambda$ playing the role of a  two-particle self-energy. The irreducible and the full vertex are connected by a Bethe-Salpeter equation.  Unlike the one-particle irreducibility, the two-particle irreducibility is ambiguous when we go beyond single-site scatterings \cite{Janis:2001ab}.  Two-particle irreducibilities  are characterized by different Bethe-Salpeter equations. Here we introduce only the Bethe-Salpeter equation in the electron-hole scattering channel
  \begin{multline}\label{eq:BS-eh}
    \Gamma_{\mathbf{k}\mathbf{k}'}(z_{1},z_{2};\mathbf{q}) 
    \\
   = \Lambda_{\mathbf{k}\mathbf{k}'}(z_{1},z_{2};\mathbf{q}) + \frac
    1N\sum_{\mathbf{k}''}
    \Lambda_{\mathbf{k}\mathbf{k}''}(z_{1},z_{2};\mathbf{q})
    G_{\mathbf{k}'' + \vecq}(z_{1}) G_{ \mathbf{k}''}(z_{2})
    \Gamma_{\mathbf{k}''\mathbf{k}'}(z_{1},z_{2};\mathbf{q})\ . 
  \end{multline}
We use this Bethe-Salpeter equation to introduce the irreducible vertex $\Lambda$ that is important for controlling consistency of approximations, more precisely whether they comply with the exact relations between one- and two-particle Green functions.    

There are no direct connections between the configurationally dependent one- and two-particle Green functions in disordered systems without inter-particle interactions. It is no longer true for the averaged Green functions. When constructing approximations one has to comply with the exact relations expressed as Ward identities that are microscopic conditions  for macroscopic conservation laws to hold.   

The basic conservation law in quantum systems is conservation of probability, that is completeness of the basis formed by the Bloch waves. A first relation between the averaged one- and two-particle Green functions follows from a simple identity for operator (matrix) multiplication  
\begin{equation}\label{eq:Ward-operator}
  \frac 1{z_1 -\widehat{H}} \ \frac 1{z_2 -\widehat{H}} = \frac 1{z_2 - z_1}
  \left[ \frac 1{z_1 -\widehat{H}} - \frac 1{z_2 -\widehat{H}}\right]\,.
\end{equation}
If we average both sides of this identity we obtain the \index{Velick\'y-Ward identity} Velick\'y-Ward identity \cite{Velicky:1969aa}
\be
  \label{eq:VW_momentum}
  \frac 1N\sum_{{\bf k}'}G^{(2)}_{{\bf k}{\bf k}'}(z_1,z_2;{\bf 0})= \frac
  1{z_2 - z_1} \left[ G({\bf k},z_1) -\ G({\bf
      k},z_2)\right]\, .
\ee
It holds provided the Bloch waves form a complete basis in the one-particle representation space. It means that the effect of the random potential is only a rotation in the Hilbert space of states of the homogeneous system.  

There is another identity connecting the irreducible one and two-particle functions, $\Sigma$ and $\Lambda$. It is a microscopic condition that guarantees validity of the macroscopic \index{continuity equation} continuity equation. By analyzing the perturbation contributions to one- and two-particle functions Vollhardt and W\"olfle proved the following \index{Volhardt-W\"olfle-Ward identity} Volhardt-W\"olfle-Ward identity \cite{Vollhardt:1980ab}
\begin{equation}
  \label{eq:VWW_identity}
  \Sigma({\bf k}_+,z_+) - \Sigma({\bf k}_-, z_-) = \frac 1N \sum_{{\bf
      k}'}\Lambda_{{\bf k}{\bf k}'}(z_+,z_-;{\bf q})\left[G({\bf
      k}'_+,z_+) - G({\bf k}'_-, z_-) \right] \ .
\end{equation}
We denoted ${\bf k}_\pm={\bf k}\pm{\bf
  q}/2$.  Identity \eqref{eq:VWW_identity} is related to
Eq.~\eqref{eq:VW_momentum}, however, the two Ward identities are identical
neither in the derivation nor in the applicability and validity domains.
The latter holds for nonzero transfer momentum $\mathbf{q}$, i.~e., for an
inhomogeneous perturbation, while the former only for $q=0$. The former is nonperturbative while the latter is proved only perturbatively.

\subsection{Feynman diagrams and multiple-occupancy corrections}

The contribution from the scatterings of the electron on the random potential can be represented diagrammatically in analogy with the \index{many-body perturbation theory} many-body perturbation theory. There is, however, an important difference between the two perturbation expansions. The former is static, contains only elastic scatterings where energy is conserved. There are no closed loops in its diagrammatic representation. The individual particles are characterized by a fixed energy, Matsubara frequency in the many-body formalism. Since the perturbation theory of random systems is static we have to introduce the so-called \index{multiple-occupancy corrections} multiple-occupancy corrections if we want to keep unrestricted summations over the lattice sites in the representations of physical quantities \cite{Yonezawa:1968aa}. We demonstrate it on Green functions.

\begin{figure}
  \begin{center}
    \includegraphics[width=16cm]{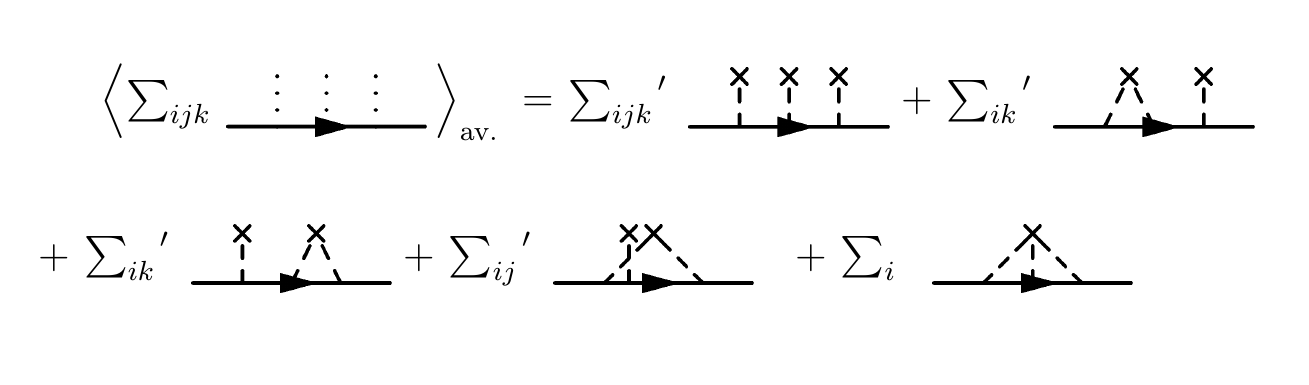}
   \caption{ Averaging of contributions up to third order of perturbation expansion in the random potential of the one-particle Green function. Prime means that the multiple sums do not contain any repeating indices, each vertex is on a different lattice site ($i\neq j\neq k$). Adapted from Ref.~\cite{Kolorenc:2004aa}.}  \label{fig:Feynman-G1-3} 
  \end{center}  
 \end{figure}

The standard diagrammatic representation of scatterings of particles on the random potential is an oriented solid line for the particle, a cross (vertex) for the lattice coordinate of the random potential and dashed lines connecting the vertex with the solid line. The number of lines staring at the vertex stands for the power of the random potential. Since the random values of the potential are independently distributed at each lattice site, we average separately each vertex of the diagrammatic expansion. When summing over the lattice sites we have to avoid repetition of the lattice indices in the multiple sums as shown in Fig.~\ref{fig:Feynman-G1-3}.    
This restriction in multiple summations makes the perturbation expansion difficult to sum and is equivalent to the excluded volume problem of  the T matrix expansion in Eq.~\eqref{eq:TMatrix-expansion}.  
 
\begin{figure}
  \begin{center}
    \includegraphics{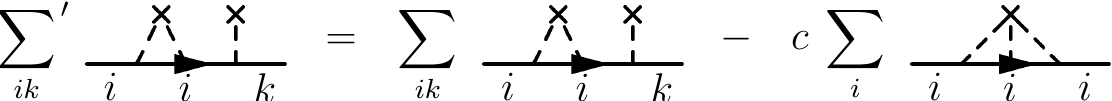}
   \caption{Transformation of restricted multiple summations to unrestricted ones with multiple-occupation corrections exemplified in third order of the perturbation expansion. Here $c$ is the concentration of the sites with the random potential. The first diagram on the right-hand side is proportional to $c^{2}$ while the second diagram only to $c$.}  \label{fig:FeynmanSC-multiple} 
  \end{center}  
 \end{figure}

 One has to transform the restricted multiple summations to unrestricted ones so that to be able to reach any nonperturbative results containing multiple scatterings.  The transformation from restricted to unrestricted sums is performed by means of \index{multiple-occupancy corrections} multiple-occupancy corrections that subtract events when any two or more lattice sites in the multiple sum are equal, as exemplified in Fig.~\ref{fig:FeynmanSC-multiple}. Counting of the multiple-occupancy corrections becomes more and more cumbersome with the increasing order of the perturbation expansion.
 
\begin{figure}
  \begin{center}
    \includegraphics{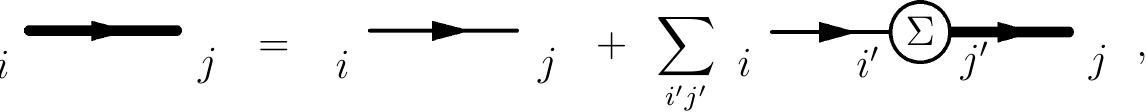}
   \caption{Diagrammatic representation of the renormalization of the one-electron propagator via the Dyson equation. Note that the sum over the primed indices is unrestricted. }  \label{fig:Feynman-Dyson} 
  \end{center}  
 \end{figure}

 Only after we have transformed the restricted sums to unrestricted ones we can introduce renormalizations of the particle lines in the  diagrammatic representation of the perturbation expansion. The renormalization of the one-particle propagator is expressed via the Dyson equation and the self-energy$\Sigma$, see Fig.~\ref{fig:Feynman-Dyson}.  All the multiple-occupancy corrections are contained in the self-energy. Its third order is diagrammatically represented as
 \be
 \includegraphics{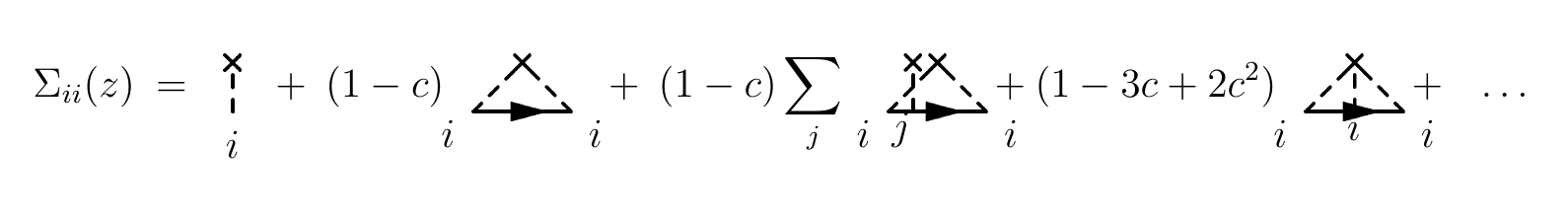}
 \ee
 It is, however, impossible to sum up the multiple-occupancy corrections to infinite order. 
 
 The diagrams with multiple-occupancy corrections offer a possibility to directly renormalize all particle lines in the diagram, that is, to replace the bare propagators with the full averaged ones. Inability of finding an analytic expression for the full sum for the self-energy to infinite order reflects the fact that we cannot construct a generating functional consisting of only the renormalized propagators, even in the local mean-field approximation, CPA. It was a breakthrough to find an analytic expression for the Soven equation in terms of the full local averaged propagator $G$  and the self-energy $\Sigma$. 
 
 %

Once we got rid of the restricted summations over lattice sites we can use the Fourier transform to momenta or wave vectors in which the Dyson equation becomes algebraic and easy to solve. The renormalization of the perturbation expansion is no longer that easy for the two-particle Green function. The averaged two-particle  Green function with three independent momenta is diagrammatically represented as
\be\label{eq:2P-GF}
\includegraphics{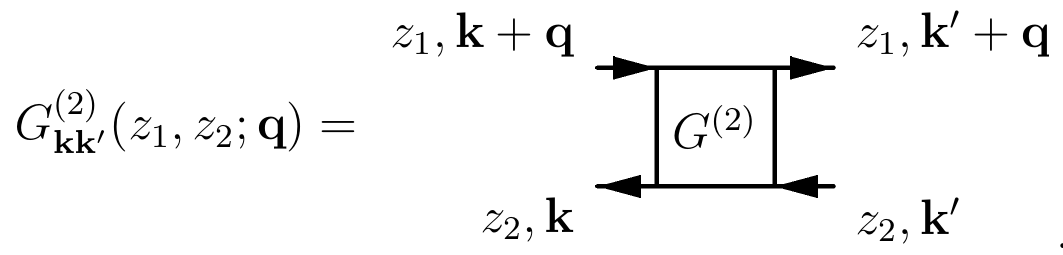}\,.
\ee
 
 The two-particle renormalization is contained in the two-particle irreducible vertex $\Lambda$  and a Bethe-Salpeter equation, a  two-particle analogy of the Dyson equation. Its diagrammatic representation reads
 \be\label{eq:2P-EH-BS}
\includegraphics{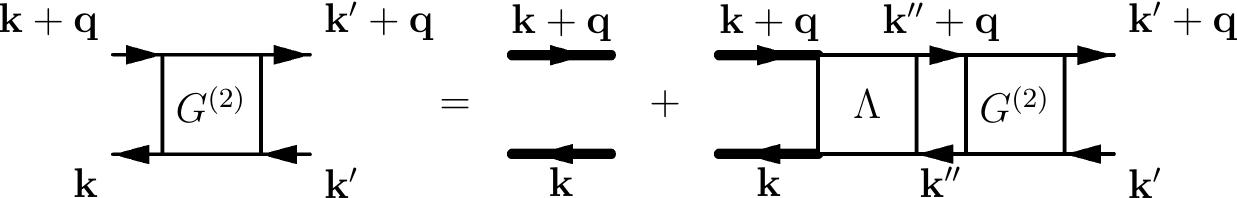}\,,
\ee
 where we sum over the double-primed momentum. 
 We introduced in Sec.~\ref{sec:GF} the full two-particle vertex which obeys an analogous Bethe-Salpeter equation where the absolute term is the irreducible vertex $\Lambda$, Eq.~\eqref{eq:BS-eh}.  Its expansion to third order of the perturbation expansion with the multiple-occupancy corrections is 
\be
 \includegraphics{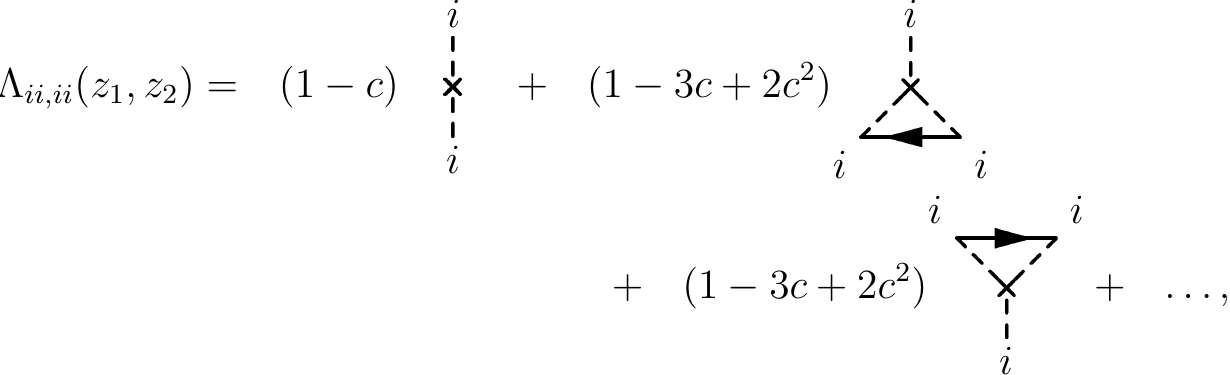}
 \ee
The particle lines can again be directly renormalized and replaced by the full averaged one-particle Green function. 
%
%
%
 
\section{Generating functional for CPA and DMFT} 
 
\subsection{Functional-integral representation of the thermodynamic potential}

The concept of the best single-site approximation can be generalized beyond quantum mechanics of disordered systems.  The best way to do so is to use a \index{functional integral} functional integral with which we can describe classical, quantum, disordered and interacting systems in a unified way. We start with a functional-integral representation of a general partition sum of an interacting and/or disordered system \cite{Janis:1999ab}
\begin{equation}
\label{Z-int}{\sf Z}\left\{ G^{(0)-1}\right\} =\int {\cal D}\varphi {\cal D}%
\varphi ^{*}\exp \left\{ -\varphi ^{*}\eta G^{(0)-1}\varphi +h^{*}\varphi
+\varphi ^{*}h+U\left[ \varphi ,\varphi ^{*}\right] \right\}\,, 
\end{equation}
where $\varphi $ and $\varphi ^{*}$ are fluctuating commuting or
anticommuting Gaussian fields, $G^{(0)-1}$ is essentially the dispersion
relation of the model represented as an inverse of the free propagator of
one-body excitations. The variable $\eta =\pm 1$ is a sign depending on
whether we deal with bosonic (commuting) or fermionic (anticommuting)
fluctuating fields, respectively. Further, $h$ is an external source and $U$
is an interaction or a random potential, i.e. a non-quadratic or inhomogeneous function of the fluctuating fields.  In Eq.~(\ref{Z-int}) we suppressed all internal degrees of freedom of the
local fluctuating fields that depend upon particular models under consideration.
The thermodynamic potential as a functional of $G^{(0)-1}$ then is 
\begin{equation}
\label{F-nsc}\Omega\left\{ G^{(0)-1}\right\} =-\beta ^{-1}\ln {\sf Z}\left\{
G^{(0)-1}\right\} \,, 
\end{equation}
where $\beta = 1/k_{B}T$.

It was the idea of Baym \cite{Baym:1962aa} to replace the functional dependence of the
thermodynamic potential on the bare propagator by a new representation with a renormalized propagator $G$. The
full propagator $G$ may be defined from the thermodynamic potential itself: 
\begin{equation}
\label{G-def}G=-\frac{\delta ^2\beta \Omega}{\delta h^{*}\delta h}=\eta \left(
\left\langle \varphi \varphi ^{*}\right\rangle -\left\langle \varphi
\right\rangle \left\langle \varphi ^{*}\right\rangle \right) =\eta \left\{ 
\frac{\delta \beta \Omega}{\delta G^{(0)-1}}-\frac{\delta \beta \Omega}{\delta h^*}\frac{
\delta \beta \Omega}{\delta h}\right\} \,.
\end{equation}
We now introduce the full propagator $G$ as a new variable into the thermodynamic potential by a substitution 
\begin{equation}
\label{S2}G^{(0)-1}=G^{-1}+\Sigma \,,
\end{equation}
where $\Sigma $ is the self-energy. We used here the Dyson equation to relate the bare and renormalized propagators. The self-energy $\Sigma$ is an accompanying variable that also enters the functional representation of the thermodynamic potential. We can treat the renormalized quantities $G$ and $\Sigma $ as independent variables in the thermodynamic potential. The new functional must, however, not depend on variations of the new variables $G$
and $\Sigma $ in order to keep the thermodynamic relations fulfilled. To
secure vanishing of variations of the thermodynamic potential with respect to $G$ and $%
\Sigma $ we have to modify the functional-integral representation, since the variation
with respect to $G^{(0)-1}$ does not vanish. We must add a contribution
being a function of only $\Sigma $ and a contribution being a function of
only $G$. If we denote them $\Omega_\Sigma $ and $\Omega_G$ we must fulfill the
following equations to keep variations of the total free energy $\Omega=\Omega_\Sigma +\Omega_G+\Omega \left\{
G^{(0)-1}\right\} $ independent of $\Sigma $ and $G$:%
$$
\frac{\delta \beta \Omega_\Sigma }{\delta \Sigma }=\frac{\delta \beta \Omega_G}{\delta
G^{-1}}=-\frac{\delta \beta \Omega}{\delta G^{(0)-1}}\,. 
$$
Using equations (\ref{G-def}) and (\ref{S2}) we easily obtain 
\begin{align}
\label{F-add}
\beta \Omega_\Sigma &=\eta\left\{
\mbox{tr}\ln \left[ G^{(0)-1}-\Sigma \right] +m^{*}\left[ G^{(0)-1}-\Sigma
\right] m\right\}, \\
\beta \Omega_G&=-\eta \left[\mbox{tr}\ln G^{-1}+m^{*}G^{-1}m\right]\, .
\end{align}
Here we had to introduce new renormalized variables 
\begin{align}
\label{m-def}m=-\frac{\delta \beta\Omega}{\delta h^{*}}\quad&,\qquad m^{*}=-\frac{\delta \beta \Omega}{\delta h}\, .
\end{align}
We use the above definitions and obtain a new representation of the thermodynamic potential 
\begin{multline}
\label{FE-full}
-\beta \Omega\left[ m;G^{-1},\Sigma \right] =-\eta
\mbox{tr}\ln \left[ G^{(0)-1}-\Sigma \right] +\eta \mbox{tr}\ln G^{-1}-\beta \Omega\left\{
h;G^{-1}+\Sigma \right\}
 \\
-m^{*}\eta\left[ G^{(0)-1}-\Sigma \right] m+m^{*}\eta G^{-1}m \,,
\end{multline}
The thermodynamic potential  $\Omega[ m;G^{-1},\Sigma]$  in Eq.~(\ref{FE-full}) is stationary (extremal) with respect to all its{\cal \ }renormalized variables $m,\Sigma ,$ and $G$. The
stationarity with respect to the variables $m,m^{*}$ leads to trivial
equations. We can, however, turn these variables dynamic if we use a
substitution in the functional integral (\ref{Z-int}) 
\begin{equation}
\label{S1}\varphi =\phi +m\,,
\end{equation}
where the new fluctuating field $\phi $ has vanishing first moment $%
\left\langle \phi \right\rangle =0$. We then obtain from Eqs.~(\ref{Z-int}), (\ref
{S1}) and (\ref{S2}) 
\begin{multline}
\label{eq:FE-mfull}
-\beta \Omega\left[ m,H;G,\Sigma \right] =-\eta
\mbox{tr}\ln \left[ G^{(0)-1}-\Sigma \right] +\eta \mbox{tr}\ln G^{-1}-\beta
F\left[ m,H;G^{-1}+\Sigma \right]   \\
 -m^{*}\eta G^{(0)-1}m+H^{*}m+m^{*}H\, .
\end{multline}
Now, the new free energy as a functional of $m,H$ and $\left[ G^{-1}+\Sigma
\right] $ reads 
\begin{align}
\label{eq:F-int}
\lefteqn{-\beta F\left[ m,H;G^{-1}+\Sigma \right] = \ln \int {\cal D}
\phi {\cal D}\phi ^{*}} 
\nonumber\\
&&\exp \left\{ -\phi ^{*}\eta
\left[G^{-1}+\Sigma \right]\phi +H^{*}\phi +\phi ^{*}H+U\left[ \phi +m,\phi
^{*}+m^{*}\right] \right\} \,,
\end{align}
where the new external sources $H$ and $H^{*}$ are new variational
variables, the Legendre conjugates to $m^{*}$ and $m$, respectively.  The
variational variables and  functions $m,H,\Sigma ,$ and $G$ are determined from
the saddle-point equations for  stationarity of $\beta \Omega{\cal \,}$: 
\begin{equation}
\label{stat eq}\frac{\delta \beta \Omega}{\delta H}=\frac{\delta \beta \Omega}{\delta m%
}=\frac{\delta \beta \Omega}{\delta G}=\frac{\delta \beta \Omega}{\delta \Sigma }=0\,.
\end{equation}
Expressions (\ref{eq:FE-mfull})-(\ref{stat eq}) are exact in any spatial
dimension for any model, classical or quantum.

The thermodynamic potential from Eq.~(\ref{eq:FE-mfull}) is not yet in the Baym form with the \index{Luttinger-Ward functional} Luttinger-Ward functional. Although it is a functional of only renormalized quantities,
diagrammatic representation of $\beta \Omega$ contains the sum of all connected
non-renormalized diagrams with the bare propagator $G^{(0)}=\left[
G^{-1}+\Sigma \right] ^{-1}$. Thermodynamic potential  (\ref{eq:FE-mfull}) is suitable for
the exact solution and for the cases where the result cannot be generated by
a sum of simple skeleton diagrams. If we have to rely on sums of classes of
particular diagrams it is more practical if we define a new functional 
\begin{align}
\label{Psi-func}
\Psi \left[ m,H;\Sigma \right] &=
\eta \mbox{tr}\ln G^{-1}-\beta \Omega\left[ m,H;G^{-1}+\Sigma \right] \,,
\end{align}
that is, due to the stationarity equations (\ref{stat eq}), independent of
the one-body propagator $G$. We have no diagrammatic representation for the
functional $\Psi .$ But if we perform a Legendre transform from $\Psi $ to a
functional of the propagator $G$ 
\begin{equation}
\label{Phi-func}\Phi \left[ m,H;G\right] =\Psi \left[ m,H;\Sigma \right]
+\eta \mbox{tr}\Sigma G \,,
\end{equation}
it will be a sum of all connected diagrams free of self-insertions, i.e.
skeleton diagrams only. Inserting Eqs.~(\ref{Psi-func}) and (\ref{Phi-func}) into
Eq.~(\ref{eq:FE-mfull}) we reveal the Baym free-energy functional %
\begin{multline}\label{FE-Baym}
-\beta \widetilde{\Omega}\left[ m,H;G\right]
= -\eta \mbox{tr}\ln \left[ G^{(0)-1}-\Sigma \right] -\eta 
\mbox{tr}\Sigma G+\Phi \left[ m,H;G\right] \\
-m^{*}\eta G^{(0)-1}m+H^{*}m+m^{*}H\,. 
\end{multline}
Both free-energy functionals (\ref{eq:FE-mfull}) and (\ref{FE-Baym}) are exact.
They are connected by a double Legendre transform (\ref{Psi-func}), (\ref{Phi-func}%
). While representation (\ref{eq:FE-mfull}) is applicable without
restrictions, the direct application of the Baym functional is restricted to
cases where we are able to find a diagrammatic representation of the
functional $\Phi $. We then speak about \index{$\Phi $-derivable approximations} $\Phi $-derivable approximations. Not all approximations are $\Phi $-derivable. The simplest example for a non- $\Phi $-derivable theory is a $0$-dimensional lattice (single site or atomic solution), including CPA and DMFT.

We considered a homogeneous system of interacting particles. It is, however, easy to extend this description to systems with randomness. If the non-quadratic term, interaction becomes random, we simply perform configurational averaging of the free energy $F\left[ m,H;G^{-1}+\Sigma \right]$  from Eq.~\eqref{eq:F-int}. The thermodynamic functional is self-averaging and hence it equals its averaged value in the thermodynamic limit.

\subsection{The limit to infinite lattice dimensions}

We have not yet made any assumption on the form of the interacting term in the functional representation of the thermodynamic potential. A special class of problems are those with a local interaction in the tight-binding representation of statistical systems in crystalline solids for which the limit to high spatial dimensions reduces the lattice to an impurity model.  

The fundamental condition in limiting the lattice models to infinite spatial dimensions is the necessity to keep the total energy of the system proportional to volume. It means that we must rescale appropriately non-local terms in the Hamiltonian.  In the case of fermions the fluctuating fields are Grassmannian variables in representation~\eqref{Z-int} and $\langle\varphi\rangle \equiv0$.    The leading-order contribution of the non-local part of the generic Hamiltonian from Eq.~\eqref{eq:Anderson-Hamiltonian},  kinetic
energy, is 
\begin{equation}
\label{E-kin}E_{kin}=-t\sum_{<ij>\sigma }\left\langle c_{i\sigma }^{\dagger
}c_{j\sigma }\right\rangle _{av}=-it\sum_{<ij>\sigma }G_{ij,\sigma
}(0^{+})\propto 2 d{\cal N}t^2  \,,
\end{equation}
where $G_{ij,\sigma }(t)=-i\left\langle {\cal T}\left[ c_{i\sigma
}(t)c_{j\sigma }^{\dagger }(0)\right] \right\rangle $ is the time-dependent
Green's function, eventually the averaged Green's function \cite{Janis:1992aa}. The scaling of
the hopping amplitude between the nearest neighbors on a hypercubic lattice with $2d$ nearest neighbors follows from Eq.~(\ref{E-kin}): 
\begin{equation}
\label{eq:hop-scal}t=t^{*}/\sqrt{2d}\,. 
\end{equation}
It was derived for the first time by Metzner and Vollhardt in Ref.~\cite{Metzner:1989aa} in the
context of the Hubbard model.

The general functional-integral representation of the thermodynamic potential with the renormalized one-electron Green function and the self-energy, Eq.~\eqref{eq:F-int}, offers a direct way to find the generating functional of the solution of the model with the local interaction/disorder in infinite dimensions, DMFT.  The scaling of the hopping term, Eq.~\eqref{eq:hop-scal}, leads to 
\begin{align}
 G&=G^{diag}\left[d^0\right]+G^{off}\left[d^{-1/2}\right],\\
      \Sigma &=\Sigma^{diag}\left[d^0\right]
+\Sigma^{off}\left[d^{-3/2}\right]\,.
\end{align}
The functional integral in Eq.~\eqref{eq:F-int} turns local in the limit $d=\infty$. It still may be a functional (infinite-dimensional) integral\index{functional integral} if the number of the  local degrees of freedom is infinite. In the case of quantum itinerant models the local degrees of freedom are spin and Matsubara frequencies. The Matsubara frequencies are coupled in the Hubbard model \cite{Janis:1992aa}. They are decoupled in the case of the Fermi gas in a random potential with the Hamiltonian from Eq.~\eqref{eq:Anderson-Hamiltonian}. The functional integral in Eq.~\eqref{eq:F-int} reduces to a product of simple integrals for individual Matsubara frequencies. The integrals can explicitly be performed and inserting  the result in Eq.~\eqref{FE-Baym} we obtain a generating functional for the  Coherent Potential Approximation  of spinless particles \cite{Janis:1989aa}
\begin{multline}\label{eq:GP-CPA}
{\cal N}^{-1}\Omega \left[ G_n,\Sigma _n\right] =-\beta
^{-1}\sum_{n=-\infty }^\infty e^{i\omega _n0^{+}}
\left\{ \int\limits_{-\infty }^\infty d\epsilon \rho _\infty
(\epsilon )\ln \left[ i\omega _n+\mu -\Sigma _n-\epsilon \right] 
\right. \\ \left.
\phantom{\int} +\left\langle \ln \left[ 1+G_{n}\left(\Sigma _n- V _i\right)\right]
\right\rangle _{av}\right\} \,. 
\end{multline}
 This representation of the CPA grand potential for a fixed chemical potential $\mu$ was  derived as the exact grand potential for the 
model of disordered electrons in $d=\infty $. Apart from the scaling of the
nearest-neighbor hopping amplitude (\ref{eq:hop-scal}) we did not use any
particular property of the perturbation theory. The limit to infinite dimensions was performed on hypercubic $d$-dimensional lattices for which the density of states in infinite dimensions reads
\be\label{eq:DOS-infty}
\rho _\infty(\epsilon )=\frac 1{\sqrt{2\pi }%
t^{*}}\exp \left\{ -\epsilon ^2/2t^{*2}\right\}\,.
\ee
The grand potential from Eq.~\eqref{eq:GP-CPA} with the density of states from Eq.~\eqref{eq:DOS-infty} is the exact solution for the Hamiltonian of the Anderson disordered model, Eq.~\eqref{eq:Anderson-Hamiltonian}, in $d=\infty$.  It serves as a good local approximation for finite-dimensional systems if the appropriate density of states is used. Notice that CPA makes sense only for lattice models with well separated nearest neighbors. It has no meaning for continuous models where multiple scatterings on a continuously spread scatterers cannot be singled out and lose on relevance.

\section{Interacting disordered electrons - Falicov-Kimball model}

\subsection{Equilibrium thermodynamic properties}

The quantum itinerant models in infinite dimensions (DMFT) can be solved analytically only if the Matsubara frequencies are decoupled and the local functional integral from  Eq.~\eqref{eq:F-int} reduces to a product of integrals for individual Matsubara frequencies. It is not the case for the Hubbard model of interacting electrons \cite{Janis:1992aa}. But  
a modification of the Hubbard Hamiltonian, the so-called
\index{Falicov-Kimball model} Falicov-Kimball model, decouples Matsubara frequencies. Its easiest spinless form
is defined by the Hamiltonian \cite{Falicov:1969aa} 
\begin{equation}
\label{FK}\widehat{H}_{FK}=-t\sum_{<ij>}c_{i}^{\dagger
}c_{j}+ \sum_i \epsilon_{i}f_i^{\dagger }f_i+\sum_ic_i^{\dagger
}c_i \left(V_{i} + Uf_i^{\dagger }f_i\right)\,. 
\end{equation}

This Hamiltonian, in comparison with the Hubbard model, loses some important
properties.  A great deal of quantum dynamics goes lost in
Hamiltonian (\ref{FK}), since its equilibrium state is not a Fermi-liquid. The Falicov-Kimball model proved, nevertheless, invaluable in the construction of an analytic mean-field theory in
strong coupling of the Hubbard-type models.

 The first exact solution of this quantum itinerant model in $d=\infty $ was derived by Brandt and Mielsch \cite{Brandt:1989aa}. The full grand potential with the
renormalized variational parameters was constructed in Ref.~\cite{Janis:1991aa}. The
spinless Falicov-Kimball Hamiltonian (\ref{FK}) does not contain any
nonlocal interaction or hybridization and hence no scaling of the coupling
term is necessary. The functional $\Omega \left\{ G_\alpha ^{-1}+\Sigma
_\alpha \right\} $ is an atomic solution the partition function of which contains
a sum over two possible states of the local electrons. The dynamic electrons
have a frequency-dependent local propagator $\left( G_{\alpha n}^{-1}+\Sigma
_{\alpha n}\right) ^{-1}$. We obtain the grand potential 
\begin{multline}\label{GP-FKexact}
\frac 2{{\cal N}}\Omega \left[ G,\Sigma \right] =-\beta^{-1} \sum_{\alpha
=\pm }\left\langle\ln \left[ 1+e^{\beta \left( \mu_{\alpha} - \varepsilon_{i} -{\cal E}_{i,-\alpha} \right) }\right]\right\rangle_{av}
-\beta ^{-1}\sum_{n=-\infty }^\infty e^{i\omega _n0^{+}}\left\{ \sum_{\alpha
=\pm }\left\langle \ln \left[ 1 
\right.\right.\right. \\ \left.\left.\left.
+\ G_{n,\alpha}\left(\Sigma _{n,\alpha} - V_{i}\right)\right]\right\rangle_{av} 
 +\int\limits_{-\infty }^\infty d\epsilon \rho
_\infty (\epsilon )\ln \left[ \left( i\omega _n+\mu_{+} -\Sigma _{n,+}\right)
\left( i\omega _n+\mu_{-} -\Sigma _{n,-}\right) -\epsilon ^2\right] \right\}\,, 
\end{multline}
with 
\begin{equation}
\label{efen}{\cal E}_{i,\alpha} =-\beta ^{-1}\sum_{n=-\infty }^\infty e^{i\omega
_n0^{+}}\ln \left(\frac{1 + G_{n,\alpha}\left(\Sigma _{n,\alpha } - V_{i} -U \right)}{1+G_{n,\alpha}\left(\Sigma _{n,\alpha} - V_{i}\right)}%
\right) \,. 
\end{equation}
We used a subscript $\alpha$ to allow for a low-temperature charge order with different sublattices. Conditions on stationarity of the grand potential (\ref{GP-FKexact}) lead to
defining equations for the variational parameters $G_{\alpha n}$ and $\Sigma
_{\alpha n}$. We obtain after a few manipulations  
\begin{subequations}
\begin{align}
\label{FK-Gdef}G_{n,\alpha} &=\int\limits_{-\infty }^\infty d\epsilon \rho
_\infty (\epsilon )\frac{\left( i\omega _n+\mu _{-\alpha} -\Sigma _{n,-\alpha}
\right) }{\left( i\omega _n+\mu_{+} -\Sigma _{n,+}\right) \left(
i\omega _n+\mu_{-} -\Sigma _{n,-}\right) -\epsilon ^2} \,,
\\ \label{FK-Sdef}
1 &=\left\langle \frac{n_{i,-\alpha}}{1+G_{n,\alpha}\left(
\Sigma _{n,\alpha } - V_{i} - U\right) } + \frac{1 - n_{i,-\alpha}}{1+G_{n,\alpha}\left(
\Sigma _{n,\alpha } - V_{i}\right) }\right\rangle_{av} \,,
\end{align}
where 
\begin{equation}
\label{FK-ndef}n_{i,\alpha} =\frac 1{1+\exp \left\{ \beta \left( \varepsilon_{i} + {\mathcal E}_{i,-\alpha}
-\mu_{\alpha} \right) \right\} } 
\end{equation}
\end{subequations}
is the averaged number of static particles in the Falicov-Kimball model. We
see that the variational parameters $G_{\alpha n}$ and $\Sigma _{\alpha n}$
now depend explicitly on the Matsubara frequencies and the thermodynamics of
the model contains a portion of quantum many-body fluctuations. The
equations of motion are algebraic and  the variational variables $G_{\alpha n}$ and $\Sigma
_{\alpha n}$ depend only on a single Matsubara frequency $\omega _n$. 

It can easily be shown that if we consider a random-alloy with a diagonal disorder where the constituent $A$ with the atomic energy $U$ has concentration $x$ and the
constituent $B$ with concentration $1-x$ has the atomic energy $0$, then
the Falicov-Kimball model in $d=\infty$ coincides with CPA of such an alloy. The Falicov-Kimball Hamiltonian defines a semiclassical model with reduced dynamical quantum fluctuations. The
dynamical fluctuations are restricted, since we have only one species of
dynamical electrons. They interact with static electrons, i.e. they are
scattered on static impurity potentials distributed in the lattice. Unlike the static disorder of the alloys  the localized electrons of the Falicov-Kimball model serving as random scatterers for the mobile electrons  are thermally equilibrated, which introduces a nontrivial thermodynamics.   The semiclassical character of the Falicov-Kimball model becomes evident from the fact that the partition function of this model can be obtained in any dimension  as a static approximation in a special  functional-integral representation \cite{Janis:1994aa}.

Equations (\ref{FK-Gdef}) and (\ref{FK-Sdef}) coincide with the well-known
Hubbard-III approximation \cite{Hubbard:1964aa} if we neglect the static
electrons and replace the density of static particles by the density of the
dynamic ones. Then equation (\ref{FK-ndef}) must be forgotten
and replaced by a sum rule. Analogy between the model of random alloys and
the Hubbard-III approximation, disclosed in Ref.~\cite{Velicky:1968aa}, led in the end of the seventies of the last century to numerous attempts to improve on the weak-coupling Hartree-Fock theory by using the ''alloy analogy'' reasoning \cite{Czycholl:1986aa}. However, mean-field theories
constructed in this way and based on the Hubbard-III approximation become 
{\em thermodynamically} {\em inconsistent }and lead to an unphysical behavior \cite{Kawabata:1972aa}. The Hubbard-III approximation was made thermodynamically consistent by adding a new variational parameter \cite{Janis:1993aa}.

\subsection{Response to external perturbations}

\begin{figure}
\begin{center}
\includegraphics{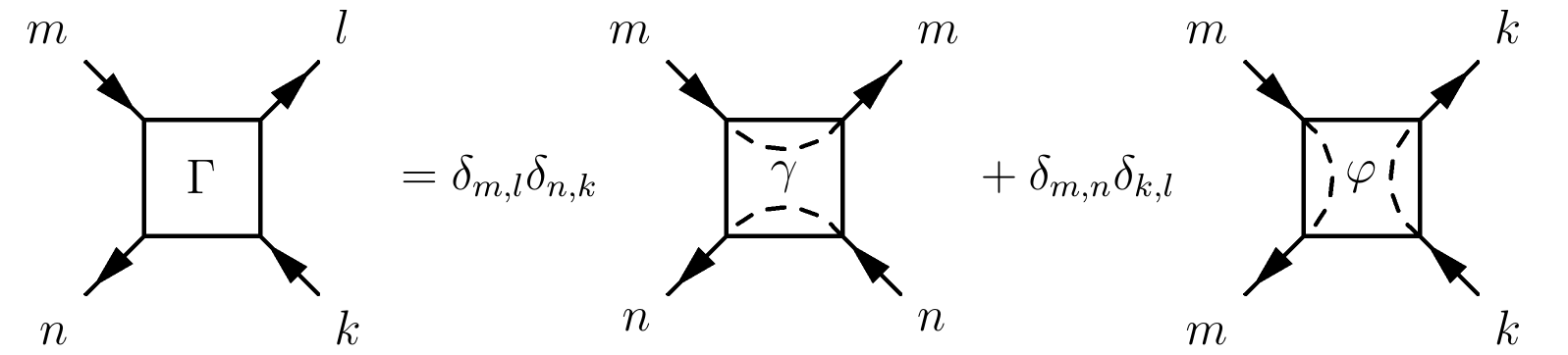}%
\caption{Graphical representation of Eq.~\eqref{eq:Gamma-local}. The
    dashed lines within the boxes indicate charge propagation from the
    incoming to the corresponding outgoing line. The vertex for the disordered Fermi gas contains only the left diagram.}  \label{fig:Gamma-local} %
\end{center}
\end{figure}

The equation for self-energy of the mobile electrons of $d=\infty$  Falicov-Kimball model for fixed densities of the local electrons resembles the Soven equation. There is, however, a significant difference when we turn to response functions describing the reaction of the equilibrium state on weak external perturbations.  The \index{response functions} response functions are derived from the two-particle Green function. The Falicov-Kimball  model, unlike the Anderson disordered model, displays a low-temperature critical behavior and a phase transition to a checker-board phase. This difference can be demonstrated on the local two-particle vertex of the  of the two solutions. In both cases, due to conservation of energy, the two-particle vertex contains only two independent variables, Matsubara frequencies. The full local vertex of the Falicov-Kimball model in $d=\infty$ measuring correlations between two conduction electrons  can be decomposed into two distinct contributions 
\begin{equation}\label{eq:Gamma-local}
  \Gamma^{MF}_{mn,kl} = \delta_{m,l} \delta_{n,k} \gamma_{m,n} +  \delta_{m,n}
  \delta_{k,l}\varphi_{m,k} \,,
\end{equation}
where the integer indices denote fermionic Matsubara frequencies. The full vertex is plotted in Fig.~\ref{fig:Gamma-local} where we indicated the way the corners of the vertices are connected via the electron line.  The CPA vertex is just its first term, $\gamma_{m,n}$, that  is relevant for transport properties (electrical conductivity). The second  vertex, $\varphi_{m,n}$, determines thermodynamic response and the low-temperature critical behavior. 

The thermodynamic vertex $\varphi_{m,n}$ can be represented via an irreducible one, $\kappa_{m,n}$, from a local Bethe-Salpeter equation 
\be
\varphi_{m,n} = \kappa_{m,n} + \frac 1\beta\sum_{l}\kappa_{m,l}G_{l}^{2}\varphi_{l,n}\,.
\ee
 The irreducible vertex satisfies a Ward identity $\kappa_{m,n} = \delta \Sigma_{m}/\delta G_{n}$ as is evident from the connection of the corners of vertex $\varphi_{m,n}$ in Fig.~\ref{fig:Gamma-local}.

It is more complicated to represent vertex $\gamma_{m,n}$ via an irreducible one. To do so and to derive the corresponding Ward identity we replicate the creation and annihilation operators and introduce
external perturbations into the thermodynamic description via a
generalized grand potential of a replicated system
$\Omega^{\nu}(\mu_1,\mu_2,\ldots\mu_{\nu};\Delta)$ with $\nu$ chemical potentials
$\mu_1,\mu_2,\ldots, \mu_{\nu}$ \cite{Janis:2005ab}. An external perturbation $\Delta$ is used to couple
different replicas and to break the initial replica independence. We
then can write
\begin{equation}
\label{eq:N-grand-potential} \Omega^{\nu}(\mu_1,\mu_2,\ldots\mu_{\nu};\Delta)
=-\frac1{\beta}\left\langle\ln\mbox{Tr}\exp
  \left\{-\beta\sum_{i,j=1}^\nu\left(\widehat{H}^{(i)}_{FKM}\delta_{ij}
    -\mu_i  \widehat{N}^{(i)}\delta_{ij} + \Delta\widehat{H}^{(ij)}
    \right)\right\} \right\rangle_{av} \,,
\end{equation}
where we assigned to each replica characterized by energy (chemical
potential) $\mu_i$ a separate Hilbert space and denoted
$\Delta\widehat{H}^{(ij)}=\sum_{kl}\Delta^{(ij)}_{kl}\widehat{c}_{k}^{(i)\dagger
}\widehat{c}^{(j)}_{l}$ an external perturbation to be set zero at the
end.  Thermodynamic potential $\Omega^{\nu}(\mu_1,\nu_2,\ldots \mu_{\nu};\Delta)$
is a generating functional for averaged products of Green functions up
to the $\nu$th order. In practice, we will use linear-response theory
with one- and two-particle Green functions, i.~e.,
$\Omega^{\nu}(\mu_1,\mu_2,\ldots \mu_{\nu};\Delta)$ is expanded up to $\Delta^2$.
Therefore it is sufficient to introduce only two replicas.

The external disturbance $\Delta$ mixes different replicas and propagators
in the replicated space are matrices in the replica indices. Since we
are interested only in the averaged two-particle functions,  we can represent 
the propagator by a two-by-two matrix 
\begin{equation}
  \label{eq:1P-matrix}   \widehat{G}^{-1}({\bf k}_1,z_1,{\bf k}_2,z_2;\Delta)
  = \begin{pmatrix}z_1-\epsilon({\bf k}_1) -\Sigma_{11}(\Delta) & \Delta -
    \Sigma_{12}(\Delta)\\ \Delta - \Sigma_{21}(\Delta) & z_2-\epsilon({\bf k}_2)
    -\Sigma_{22}(\Delta) 
    \end{pmatrix} \,,
\end{equation}
where $\epsilon({\bf k})$ is the lattice dispersion relation and the
self-energy elements $\Sigma_{ab}$ generally depend on both energies
$z_1,z_2$. 

The local  two-particle vertex is a solution of a Bethe-Salpeter equation with an irreducible two-particle vertex $\lambda$ and local propagators. We easily find  that the Bethe-Salpeter equation in the mean-field approximation reduces to an algebraic one
\begin{equation}\label{eq:local-vertex}
\gamma(z_1,z_2)=\frac{\lambda(z_1,z_2)}{1-\lambda(z_1,z_2)G(z_1)G(z_2)}
\,.
\end{equation}
The irreducible vertex $\lambda$ is determined  in equilibrium ($\Delta=0$) from an equation consistent with the Ward identity, Eq.~\eqref{eq:VWW_identity},
\begin{multline}\label{eq:2IP-vertex}
  \lambda_{m,n}   =\frac 1{G_{m}G_{n}}\left\{1- 
  \left\langle \left[\frac{n_{i,-\alpha}}{1+G_{n,\alpha}\left(
\Sigma _{n,\alpha } - V_{i} - U\right) } + \frac{1 - n_{i,-\alpha}}{1+G_{n,\alpha}\left(
\Sigma _{n,\alpha } - V_{i}\right) }\right]  
  \right.\right. \\ \left.\left.
   \times\left[ \frac{n_{i,-\alpha}}{1+G_{n,\alpha}\left(
\Sigma _{n,\alpha } - V_{i} - U\right) } + \frac{1 - n_{i,-\alpha}}{1+G_{n,\alpha}\left(
\Sigma _{n,\alpha } - V_{i}\right) }\right] \right\rangle^{-1}_{av}\right\} \ .
\end{multline}
We can easily verify that this equation coincides with the CPA
solution for the irreducible vertex $\lambda(z_1,z_2)$ \cite{Velicky:1969aa,Janis:2003aa}.

\section{Transport properties within CPA}

\subsection{Non-local two-particle vertex and electrical conductivity}

There is no ambiguity in the mean-field construction of local one- and
two-particle functions. But a mean-field treatment has a physical
relevance only if it is able to produce nonlocal correlation
functions, the long-range fluctuations of which may significantly
influence the thermodynamic and dynamical behavior.  There is not,
however, a unique way how to generate the two-particle vertex with
non-local contributions within the local (mean-field) approach. The
simplest and most straightforward way is to use the Bethe-Salpeter
equation with the CPA irreducible vertex $\lambda$,
Eq.~\eqref{eq:2IP-vertex}, and to replace the product of the local propagators with
a convolution of the full nonlocal one-electron propagators $G(\mathbf{k},z)$. Such a
Bethe-Salpeter equation then remains algebraic in momentum representation
and results in a two-particle vertex with only one transfer momentum.
We obtain
\begin{equation}\label{eq:CPA-vertex}
\Gamma^\pm(z_1,z_2;\mathbf{q}^\pm)=\frac{\lambda(z_1,z_2)}{1-
\lambda(z_1,z_2) \chi^\pm(z_1,z_2;\mathbf{q}^\pm)} \,,
\end{equation}
where we denoted the two-particle bubble
\begin{equation}\label{eq:bubble}
  \chi^{\pm}(z_1,z_2;{\bf q})=\frac 1N\sum_{\bf k} G({\bf k},z_1)G({\bf
    q}\pm{\bf k},z_2)\,.
\end{equation}
The ambiguity in this definition of the full mean-field vertex is in
the type of nonlocal multiple scatterings we include into the
Bethe-Salpeter equation. They are here denoted by the superscript
$\pm$. The plus sign corresponds to multiple scatterings of
the electron-hole pairs, while the minus sign to the pairs of electrons.
In case of elastic scatterings the electron-hole and electron-electron
bubbles produce numerically the same result. However, the difference
between the two types of pair scatterings lies in the respective
transfer momentum $\mathbf{q}^\pm$.  Using the notation for momenta in  the
two-particle Green function from Eq.~\eqref{eq:2P-GF} we have
$\mathbf{q}^+ =\mathbf{q}$ for the electron-hole pair scatterings and $\mathbf{q}^- = \mathbf{q} + \mathbf{k} + \mathbf{k}'$ for scatterings of two electrons. The \index{nonlocal vertex} nonlocal vertex in CPA is that from from Eq.~\eqref{eq:CPA-vertex} and the electron-hole bubble with $\vecq^{+}$.  We discuss more this ambiguity in the next subsection.
 
We now turn our attention to the \index{electrical conductivity} electrical conductivity. Using the  Kubo formula we obtain a simple representation of the longitudinal conductivity at zero temperature \cite{Janis:2010aa}
\begin{equation}\label{eq:Conductivity-zeroT}
  \sigma_{\alpha\alpha} = \frac{e^2}{2\pi N^2}\sum_{\mathbf{k}\mathbf{k}'}
  v_\alpha(\mathbf{k}) v_\alpha(\mathbf{k}')\left[ G^{AR}_{\mathbf{k}\mathbf{k}'}
    - \Re G^{RR}_{\mathbf{k}\mathbf{k}'}\right] \,,
\end{equation}
with the values of the two-particle Green function at the Fermi
energy. We used an abbreviation $G^{AR}_{\mathbf{k}\mathbf{k}'} =
G^{AR}_{\mathbf{k}\mathbf{k}'}(0,0;\mathbf{0})$ and 
\begin{align*}
G^{AR}_{\mathbf{k}\mathbf{k}'}(\omega,\omega';\vecq) &=
G^{\{2\}}_{\mathbf{k} \mathbf{k}'}(\omega -i0^+, \omega' +
i0^+;\mathbf{q}),\\
G^{RR}_{\mathbf{k}\mathbf{k}'}(\omega,\omega';\vecq) &=
G^{\{2\}}_{\mathbf{k} \mathbf{k}'}(\omega + i0^+, \omega' +
i0^+;\mathbf{q})\ .
 \end{align*}
We decompose the conductivity tensor into two parts by replacing the two-particle Green function by representation~\eqref{eq:G2-Gamma} with the two-particle vertex $\Gamma$. We then have a sum
\begin{equation}\label{eq:Conductivity-decomposition}
  \sigma_{\alpha\alpha} = \frac{e^2}{\pi
    N}\sum_{\mathbf{k}}\left|v_\alpha(\mathbf{k}) \right|^2 \left|\Im
G^R(\mathbf{k}) \right|^2 + \Delta\sigma_{\alpha\alpha} \,,
\end{equation}
where the first term is the standard one-electron or Drude conductivity at zero
temperature. The genuine two-particle contribution is called a vertex
correction and is proportional to the appropriate matrix element of
the two-particle vertex that, at zero temperature, reads
\begin{equation}\label{eq:Conductivity-correction}
  \Delta\sigma_{\alpha\alpha} = \frac{e^2}{2\pi
    N^2}\sum_{\mathbf{k}\mathbf{k}'} v_\alpha(\mathbf{k})
  v_\alpha(\mathbf{k}')\left\{
    \left|G^R_\mathbf{k}\right|^2\Delta\Gamma^{AR}_{\mathbf{k}\mathbf{k}'}
    \left| G^R_{\mathbf{k}'}\right|^2 
   - \Re\left[\left(G^R_\mathbf{k}\right)^2
      \Delta\Gamma^{RR}_{\mathbf{k}\mathbf{k}'}
      \left(G^R_{\mathbf{k}'}\right)^2 \right] \right\}\
  . \end{equation}
It is not the full two-particle vertex that is
important for the electrical conductivity, but only its odd part
$\Delta\Gamma$. That is, only the part of the vertex function depending on the bipartite lattices on odd powers of the fermionic momenta $\mathbf{k}$ and $\mathbf{k}'$ contributes to the electrical conductivity. Hence, CPA does not contain vertex corrections to the electrical conductivity, since the two-particle vertex $\Gamma^{CPA}_{\veck\veck'}(\omega,\omega';\vecq) = \Gamma(\omega, \omega';\vecq)$ does not depend on the incoming fermionic momenta $\veck,\veck'$.

\subsection{Gauge invariance and electron-hole symmetry}

Electrical conductivity is a form of a response of the charged system to an electromagnetic perturbation. An important feature of the interaction of the charged system with an electromagnetic field is gauge invariance that must be guaranteed in the response functions. There are two fundamental response functions to the electromagnetic field, one based on the current-current and the other on the density-density correlation functions. The former is used in the Kubo formula for the electrical conductivity, Eq.~\eqref{eq:Conductivity-zeroT}, and the latter for determination of \index{charge diffusion} charge diffusion.    

There is a number of more or less heuristic arguments in the literature that relate the density response with the conductivity \cite{Rammer:1998aa}. They use macroscopic \index{gauge invariance} gauge invariance and charge conservation for particles exposed to an electromagnetic field. A microscopic quantum derivation was presented in Ref.~\cite{Janis:2003aa}. 

Gauge invariance is used to relate the external scalar potential with the electric field $\mathbf{E}
= -\boldsymbol{\nabla}\varphi$.  The current density generated by a harmonic
external field then is
\begin{equation}
  \label{eq:induced_current}
  \mathbf{j}({\bf q},\omega) = \boldsymbol{\sigma}({\bf q},\omega)\cdot
  \mathbf{E} ({\bf q},\omega) = -  i \boldsymbol{\sigma}({\bf
q},\omega)\cdot   {\bf q}\ \varphi({\bf q},\omega) \,,
\end{equation}
where $\boldsymbol{\sigma}({\bf q},\omega)$ denotes the tensor of the
electrical conductivity.  Charge conservation is expressed by a continuity
equation. In equilibrium we can use the operator form of the \index{continuity equation}continuity equation that follows from the Heisenberg equations of motion for the current and density operators. For
Hamiltonians with a quadratic dispersion relation we have
\begin{equation}
  \label{eq:continuity_equation}
   e \partial_t\widehat{n}(\mathbf{x},t) + \boldsymbol{\nabla}\cdot
  \widehat{\mathbf{j}}(\mathbf{x},t) = 0 \ .
\end{equation}
Energy-momentum representation of the continuity equation in the
ground-state solution is
\begin{equation}
  \label{eq:continuity_Fourier}
  - i\omega e\delta n({\bf q},\omega) +i \mathbf{q}\cdot\mathbf{j}({\bf
  q},\omega) = 0 \ .
\end{equation}
We have to use a density variation of the equilibrium density, i.~e.,
the externally induced density $\delta n({\bf q},\omega) = n({\bf
  q},\omega) -n_0$ in the continuity equation with the averaged values of
the operators.  From the above equations and for linear response $\delta
n({\bf q},\omega) = -e\chi({\bf q},\omega) \varphi({\bf q},\omega)$ we
obtain in the isotropic case
\begin{equation}
  \label{eq:Einstein_general}
  \sigma = \lim_{\omega\to 0}\lim_{q\to 0}\frac{-ie^2\omega}{q^2} \chi({\bf q},\omega) \,,
\end{equation}
where at zero temperature 
\begin{multline}
\chi(\vecq,\omega) 
= -\left\{\int_{-\omega}^{0}\frac{dx}{2\pi i}\left\langle G^{AR}_{\veck\veck'}(x,x + \omega;\vecq) - G^{RR}_{\veck\veck'}(x,x + \omega;\vecq) \right\rangle_{\veck,\veck'} 
\right. \\ \left.
- \int_{-\infty}^{0}\frac{dx}{\pi}\Im\left\langle G^{RR}_{\veck\veck'}(x,x + \omega;\vecq)\right\rangle_{\veck,\veck'}\right\}
\end{multline}
is the density response function. Relation~\eqref{eq:Einstein_general} is often taken
as granted for the whole range of the disorder strength and used for
the definition of the zero-temperature conductivity when describing
the Anderson localization transition \cite{Vollhardt:1992aa,Belitz:1994aa}.

Relation~\eqref{eq:Einstein_general} between the static optical conductivity and the density response holds if the latter function displays the so-called diffusion pole. One can prove by using the Vollhardt-W\"olfle-Ward identity, Eq.~\eqref{eq:VWW_identity}, that in the limit $q\to0$ and $\omega\to 0$  \cite{Janis:2003aa}
\be\label{eq:chi-omega}
\chi(\vecq,\omega) \doteq \frac{Dn_{F} q^{2}}{-i \omega + Dq^{2}} \,,
\ee
where $D$ is the static diffusion constant.  Inserting Eq.~\eqref{eq:chi-omega} into Eq.~\eqref{eq:Einstein_general} we end up with the \index{Einstein relation} Einstein relation between the diffusion constant and conductivity   $\sigma = e^2 n_F D $, where $n_{F}$ is the electron density at the Fermi energy. This relation holds in CPA with the Drude conductivity  \cite{Janis:2003aa}.

\begin{figure}
\begin{center}
\includegraphics{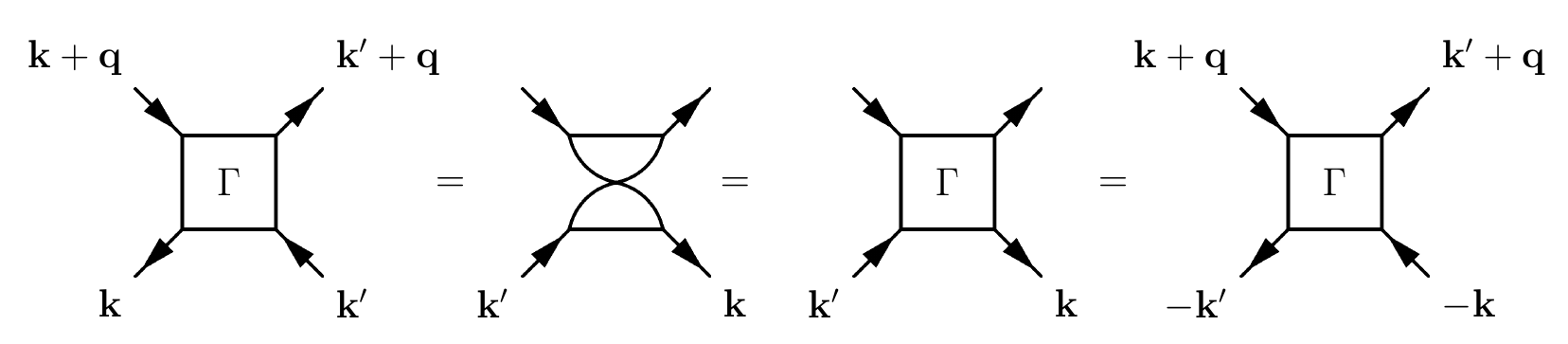}%
\caption{Graphical representation of time-reversal symmetry of the two-particle vertex when the lower electron line is reversed. }  \label{fig:Gamma-Equation} %
\end{center}
\end{figure}

The coherent potential approximation delivers good results for the one-particle quantities but it fails to take into account back-scatterings that are responsible for the vertex corrections in the electrical conductivity.  It also fails to maintain \index{electron-hole symmetry} electron-hole symmetry at the level of two-particle functions. Electron-hole symmetry or equivalently time-reversal is an important
feature of electron systems without spin- and orbital-dependent
scatterings. According to this invariance the physical (measurable)
results should not depend on the orientation of the electron propagators. Changing the orientation of the electron line is equivalent to the spatial reflection in the momentum space. The electron-hole symmetry for one- and two-particle propagators says
\begin{subequations}\label{eq:EH}
\begin{align}\label{eq:EH-GF}
  G(\mathbf{k},z)& = G(-\mathbf{k},z)\ , \\
  \Gamma_{\mathbf{k}\mathbf{k}'}(z_+,z_-;\mathbf{q})& =
  \Gamma_{\mathbf{k}\mathbf{k}'}(z_+,z_-;- \vecq - \veck - \veck')
   = \Gamma_{-\mathbf{k}'-\mathbf{k}}(z_+,z_-; \vecq + \veck + \veck')\ .
\end{align}
\end{subequations}
The spatial reflection was applied only to one fermion propagator in the two-particle vertex, the upper in the first equality and the lower in the second. The latter transformation is graphically represented in Fig.~\ref{fig:Gamma-Equation}. This is an exact relation which is, however, broken in CPA as discussed in Refs.~\cite{Janis:2005aa,Janis:2005ab}. One has to go beyond the local mean-field approximation to recover this deficiency. 

\section{Beyond CPA}


\subsection{Vertex corrections to the electrical conductivity}

The dynamical mean-field theory contains all single-site scatterings. To go beyond, one has to avoid repetition of lattice indices as in the expansion of the T matrix in Eq.~\eqref{eq:TMatrix-expansion}.    The bare expansion parameter is then the off-diagonal one-electron propagator from the mean-field theory. It is 
\begin{equation}\label{eq:G-off}
  \bar{G}(\mathbf{k},\zeta) = \frac 1{\zeta - \epsilon(\mathbf{k})} - \int
  \frac{d\epsilon \rho(\epsilon)}{\zeta -\epsilon} \,,
\end{equation}
where we denoted $\zeta= z - \Sigma(z)$ and the local self-energy
$\Sigma(z)$ is that of the mean-field solution. The off-diagonal two-particle bubble describes the simplest non-local contribution and it is convolution of the off-diagonal one-electron propagators. We have
\begin{equation}\label{eq:chi-bar}
  \bar{\chi}(\zeta,\zeta';\mathbf{q}) = \frac 1N\sum_{\mathbf{k}}
  \bar{G}(\mathbf{k},\zeta) \bar{G}(\mathbf{k} + \mathbf{q},\zeta') = \chi(\zeta,\zeta';\mathbf{q}) - G(\zeta) G(\zeta')\,, 
\end{equation}
where $\chi(\zeta,\zeta';\mathbf{q})$ is the full two-particle bubble.
The frequency indices are external parameters and we suppress them
when they are not necessary to specify the particular type of the one- or
two-electron propagators.

The asymptotic limit of the full two-particle vertex in high spatial
dimensions contains beyond the local mean-field vertex $\gamma$ also
non-local contributions from the electron-hole and electron-electron
ladders. They are different in their off-diagonal part \cite{Janis:2005aa,Janis:2005ab}. The asymptotic two-particle vertex consistent with the electron-hole symmetry can in the leading order be represented as
\begin{equation}\label{eq:Gamma-high_d}
  \Gamma_{\mathbf{k}\mathbf{k}'}(\mathbf{q}) = \gamma\left[1 + \gamma
    \left(\frac{\bar{\chi}(\mathbf{q})}{1 - \gamma\bar{\chi}(\mathbf{q})}
      + \frac{\bar{\chi}(\mathbf{Q})}{1 - \gamma\bar{\chi}(\mathbf{Q})}\right)\right] = \gamma + \Delta\Gamma_{\veck\veck'}(\vecq) \,,
\end{equation}
where we denoted $\mathbf{Q} = \mathbf{q} + \mathbf{k} + \mathbf{k}'$
the momentum conserved in the electron-electron channel. Notice that
the contribution from the electron-hole channel with
$\bar{\chi}(\mathbf{q})$ is part of the two-particle vertex from CPA and can be derived from the  Velick\'y-Ward identity \cite{Velicky:1969aa}. The two-particle vertex from CPA
does not carry the full $1/d$ correction to the local vertex and moreover
it is not electron-hole symmetric on the two-particle
level \cite{Janis:2005aa}. 

A consistent extension of the local mean-field
two-particle vertex must contain both non-local contributions
from the electron-hole and  electron-electron channels as given in
Eq.~\eqref{eq:Gamma-high_d}. It is not appropriate to calculate the electrical conductivity from the decomposition in Eq.~\eqref{eq:Conductivity-decomposition}, since the vertex corrections can overweigh the Drude term and the conductivity may get negative \cite{Pokorny:2013aa}. The actual expansion around CPA should be done for the electron-hole irreducible vertex $\Lambda$ from the Bethe-Salpeter equation~\eqref{eq:BS-eh}. Actually, only the vertex correction  $\overline{\Lambda}= \Lambda - \lambda$ is the object of the perturbation expansion around CPA. The  two-particle Green function with only the off-diagonal part of the vertex corrections can  be represented  as \cite{Pokorny:2013aa}
\begin{equation}\label{eq:BS-G2-bar}
 \overline{G}^{ab}_{\mathbf{k}\mathbf{k}'}(\mathbf{q}) = G^{b}_{\mathbf{k} + \mathbf{q}} \left[\delta(\mathbf{k} -\mathbf{k}') + \overline{G}^{a}_{\mathbf{k}} \Gamma^{ab}_{\mathbf{k}\mathbf{k}'}(\mathbf{q})  \overline{G}^{b}_{\mathbf{k}' + \mathbf{q}}   \right] G^{a}_{\mathbf{k}'} \,,
\end{equation}
where superscripts $a,b$ stand for $R,A$ where appropriate.
It is a solution of a Bethe-Salpeter equation with the irreducible vertex $\bar{\Lambda}^{ab}$
\begin{equation}\label{eq:BS-G2-bar_{symmetric}}
\overline{G}^{ab}_{\mathbf{k}\mathbf{k}'}(\mathbf{q}) = G^b_{\mathbf{k} + \mathbf{q}}\left[ 1 - \widehat{\overline{\Lambda}}\phantom{}^{ab}(\mathbf{q}) \star\right]^{-1}_{\mathbf{k}\mathbf{k}'} G^{a}_{\mathbf{k}'}\,.
\end{equation}

 We use the constrained two-electron Green function $\overline{G}^{ab}$ in the calculation of the electrical conductivity.  From the  Bethe-Salpeter equation~\eqref{eq:BS-G2-bar_{symmetric}}  we straightforwardly obtain
\begin{equation}\label{eq:sigma-bar-nonperturbative}
\sigma_{\alpha\beta} = \frac{e^{2}}{2\pi N^{2}}\sum_{\mathbf{k}\mathbf{k}'} v_{\alpha}(\mathbf{k})\left\{ G^{A}_{\mathbf{k}}\left[1 - \widehat{\overline{\Lambda}}\phantom{}^{RA} \star\right]^{-1}_{\mathbf{k}\mathbf{k}'} G^{R}_{\mathbf{k}'} 
- \Re \left( G^{R}_{\mathbf{k}}\left[1 - \widehat{\overline{\Lambda}}\phantom{}^{RR}\star \right]^{-1}_{\mathbf{k}\mathbf{k}'} G^{R}_{\mathbf{k}'}\right)\right\} v_{\beta}(\mathbf{k}'
)\ .\end{equation}
The expansion around CPA with the off-diagonal propagators will now be applied on the irreducible vertex $\bar{\Lambda}$ in the above equation accompanied by a non-perturbative matrix inversion so that not to break non-negativity of the conductivity \cite{Pokorny:2013aa}. The leading order contribution to vertex $\bar{\Lambda}^{ab}$  in high spatial dimensions is 
\begin{equation}\label{eq:Lambda-bar-eh}
\overline{\Lambda}^{ab}_{\mathbf{k}\mathbf{k}'}(\mathbf{q}) = \gamma^{ab} \left[1 + \frac{\gamma^{ab} \overline{\chi}^{ab}(\mathbf{k} + \mathbf{k}'+ \mathbf{q})}{1 - \gamma^{ab} \overline{\chi}^{ab}(\mathbf{k} + \mathbf{k}'+ \mathbf{q})} \right]\,.
\end{equation}
 
The off-diagonal propagator $\overline{G}$ is the  fundamental parameter in the expansion around the mean-field limit.   It prevents from double counting of multiple local scatterings from the mean-field solution and it also makes the calculation of corrections to the mean-field result numerically more stable.  It is preferable to use the full local mean-field vertex $\gamma^{ab}$ instead of the irreducible one, $\lambda^{ab}$, in all formulas of the \index{expansion around mean field} expansion around mean field, since  the latter contains a pole in the $RR$ ($AA$) channel that is compensated in the perturbation expansion with the former vertex. Notice that the leading-order vertex corrections calculated from the expansion of the right-hand side of Eq.~\eqref{eq:sigma-bar-nonperturbative} coincide with the leading corrections to the mean-field conductivity derived  in Ref.~\cite{Janis:2010aa}.

\subsection{Making expansions beyond local approximations conserving}

The problem of the perturbation expansions is that they  break exact relations. Sometimes the deviations from the exact relations are not drastic and do not qualitatively alter the physical behavior. It is not the case for disordered systems. The Einstein relation, Eq.~\eqref{eq:Einstein_general},  between the diffusion constant and the electrical conductivity is of fundamental importance. It holds, however, only if the Vollhardt-W\"ofle-Ward identity, Eq.~\eqref{eq:VWW_identity}, is obeyed. Each perturbation expansion for two-particle quantities breaks this identity in that a causal self-energy $\Sigma$ cannot be made compatible with a given approximate two-particle irreducible vertex $\Lambda$ by fulfilling  Eq.~\eqref{eq:VWW_identity} \cite{Janis:2004aa}.  Approximations where Eq.~\eqref{eq:VWW_identity} is broken are no longer conserving and the validity of continuity equation~\eqref{eq:continuity_equation} cannot be guaranteed.

To make the perturbation theory for two-particle functions consistent and conserving  we first reconcile the two-particle irreducible vertex in the diagrammatic expansion with the one-electron self-energy  via the Ward identity in the best possible way. The full dynamical Ward identity can neither be used to determine the one-particle self-energy from the two-particle irreducible vertex nor vice versa, since the vertex contains more information than the self-energy. The Ward identity poses a restriction on the form of the two-particle irreducible vertex and generally serves only as a consistency check and a guarantee that the macroscopic conservation laws are obeyed. Ward identity~\eqref{eq:VWW_identity} for $\omega=0$ and $q=0$ can nevertheless be used to determine the imaginary part of the self-energy from the electron-hole irreducible vertex   via   
\begin{subequations}\label{eq:KK-relations}
\begin{align}\label{eq:KK-imaginary}
\Im \Sigma^{R}_{\mathbf{k}}(E) & = \frac 1N\sum_{\mathbf{k}'}\Lambda^{RA}_{\mathbf{k}\mathbf{k}'}(E;0,\mathbf{0}) \Im G^{R}_{\mathbf{k}'}(E) \,,
\end{align}
since both sides of this identity contain the same number of degrees of freedom and the equation for the imaginary part of the self-energy  can consistently be resolved for each energy $E$ and momentum $\veck$. The corresponding real part of the self-energy is then found from the \index{Kramers-Kr\"onig relation} Kramers-Kr\"onig relation
\begin{align}\label{eq:KK-real}
\Re \Sigma^{R}_{\mathbf{k}}(E) & = \Sigma_{\infty} + P\int_{-\infty}^{\infty} \frac{d\omega}{\pi} \frac{\Im\Sigma^{R}_{\mathbf{k}}(\omega)}{\omega - E}
\end{align}
\end{subequations}
that ensures analyticity and causality of the self-energy in the plane of complex energies beyond the real axis. 

Since we know that the full dynamical Ward identity cannot be fulfilled by the irreducible vertex from the perturbation expansion we introduce a new physical irreducible vertex that we denote $L$. It will be connected with vertex $\Lambda$ from the perturbation theory but will be made to obey the full Ward identity, that is,    
\begin{equation}\label{eq:WI-VW}
\Delta \Sigma_{\mathbf{k}}^{RA}(E;\omega,\mathbf{q})   = \frac 1{N}\sum_{\mathbf{k}'} L^{RA}_{\mathbf{k}_{+},\mathbf{k}'_{+}}(E, \omega;\mathbf{q}) \Delta G_{\mathbf{k}'}^{RA}(E;\omega,\mathbf{q}) 
\end{equation}
holds. Here we introduced 
$
\Delta G_{\mathbf{k}}^{RA}(E;\omega,\mathbf{q})  = G^{R}_{\mathbf{k}_{+}}(E_{+}) - G^{A}_{\mathbf{k}_{-}}(E_{-}) $
and
$\Delta \Sigma_{\mathbf{k}}^{RA}(E;\omega,\mathbf{q})  = \Sigma^{R}_{\mathbf{k}_{+}}(E_{+}) - \Sigma^{A}_{\mathbf{k}_{-}}(E_{-}) $. We denoted $\veck_{\pm}=\veck \pm \vecq/2, E_{\pm} = E \pm \omega/2$.

Vertex $L^{RA}_{\mathbf{k}_{+},\mathbf{k}'_{+}}(E, \omega;\mathbf{q}) $ is not directly accessible in diagrammatic approximations.  The two-particle vertex functions in the perturbation expansion are represented by classes of diagrams with sums over momenta in the whole two-particle Hilbert space. The output of the diagrammatic expansion is an irreducible vertex $\Lambda_{\veck\veck'}(E;\omega,\vecq)$ that does not \textit{generically} comply with Ward identity~\eqref{eq:WI-VW} if the self-energy $\Sigma^{R/A}_{\veck}(E)$ is non-local, that is, depends on momentum $\veck$. Vertex corrections that take into account the impact of the Ward  identity on the irreducible vertex for the given self-energy then must be introduced beyond the standard diagrammatic approach to make the theory conserving.  We therefore distinguish the physical vertex  $\Gamma^{RA}$ obeying the Bethe-Salpeter equation with the irreducible vertex $L^{RA}$  from vertex $\widetilde{\Gamma}^{RA}$ determined from the  perturbative vertex $\Lambda^{RA}$  via the corresponding Bethe-Salpeter equation~\eqref{eq:BS-eh}. 

 We can make the approximations for the  two-particle vertex $\Lambda_{\mathbf{k}\mathbf{k}'}(E;\omega,\mathbf{q})$ conserving by appropriately correcting its action in the momentum space. It is namely sufficient to correctly replace the values of vertex $\Lambda$ on a subspace on which its action is already predefined by the self-energy via the Ward identity. For this purpose we introduce a new correcting function measuring the deviation of the given vertex $\Lambda^{RA}$ from the Ward identity \cite{Janis:2016aa}
\begin{equation}\label{eq:R-def}
R_{\mathbf{k}}(E;\omega,\mathbf{q})  = \frac 1N\sum_{\mathbf{k}'}\Lambda^{RA}_{\mathbf{k}_{+}\mathbf{k}'_{+}}(E;\omega,\mathbf{q}) \Delta G_{\mathbf{k}'}(E;\omega,\mathbf{q})
 - \Delta \Sigma_{\mathbf{k}}(E;\omega,\mathbf{q}) \ .
\end{equation}
This function vanishes in the metallic phase for $\omega=0$ and $q=0$ due to the definition of the self-energy, Eq.~\eqref{eq:KK-relations}. It is identically zero if vertex $\Lambda_{\mathbf{k}\mathbf{k}'}(E;\omega,\mathbf{q})$ obeys the Ward identity. With the aid of function $R_{\mathbf{k}}(E;\omega,\mathbf{q})$ we construct a conserving electron-hole irreducible vertex \cite{Janis:2016aa}
\begin{multline}\label{eq:L-Lambda}
L^{RA}_{\mathbf{k}_{+}\mathbf{k}'_{+}}(E;\omega,\mathbf{q}) = \Lambda^{RA}_{\mathbf{k}_{+}\mathbf{k}'_{+}}(E;\omega,\mathbf{q})  - \frac 1{\langle\Delta G(E;\omega,\mathbf{q})^{2}\rangle}
\left[\Delta G_{\mathbf{k}}(E;\omega,\mathbf{q}) R_{\mathbf{k}'}(E;\omega,\mathbf{q}) 
\phantom{\frac 12}\right. \\ \left. 
+ R_{\mathbf{k}}(E;\omega,\mathbf{q}) \Delta G_{\mathbf{k}'}(E;\omega,\mathbf{q}) \phantom{\frac 12}
 - \frac{\Delta G_{\mathbf{k}}(E;\omega,\mathbf{q}) \Delta G_{\mathbf{k}'}(E;\omega,\mathbf{q})}{\left\langle\Delta G(E;\omega,\mathbf{q})^{2}\right\rangle}
 \left\langle R(E;\omega,\mathbf{q}) \Delta G(E;\omega,\mathbf{q})\right\rangle\right]
\end{multline}
that is manifestly compliant with the full Ward identity, Eq.~\eqref{eq:WI-VW},  for arbitrary $\omega$ and $\vecq$. We abbreviated $\langle\Delta G(E;\omega,\mathbf{q})^{2}\rangle = N^{-1}\sum_{\mathbf{k}} \Delta G_{\mathbf{k}}(E;\omega,\mathbf{q})^{2}$ and\\  $\left\langle R(E;\omega,\mathbf{q}) \Delta G(E;\omega,\mathbf{q})\right\rangle = N^{-1}\sum_{\mathbf{k}}R_{\mathbf{k}}(E;\omega,\mathbf{q}) \Delta G_{\mathbf{k}}(E;\omega,\mathbf{q})$. 

Function $L^{RA}$ is the desired physical irreducible vertex to be  used in determining the physical vertex $\Gamma^{RA}$ from which all relevant  macroscopic quantities will be calculated. The \index{conserving vertex} conserving vertex  $\Gamma_{\veck\veck'}(E;\omega,\mathbf{q})$,  determined from the Bethe-Salpeter equation with the conserving  irreducible vertex $L_{\veck\veck'}(E;\omega,\mathbf{q})$, generally differs from vertex $\widetilde{\Gamma}(E;\omega,\mathbf{q})$ obtained from the perturbative vertex  $\Lambda_{\veck\veck'}(E;\omega,\mathbf{q})$.  The two vertices are equal only when the difference function $R_{\mathbf{k}}(E;\omega,\mathbf{q})$ vanishes. Our construction guarantees that it happens for $\omega=0$ and $q=0$, that is,
\begin{equation}
\widetilde{\Gamma}^{RA}_{\mathbf{k}\mathbf{k}'}(E;0,\mathbf{0}) = \Gamma^{RA}_{\mathbf{k}\mathbf{k}'}(E;0,\mathbf{0})  \ ,
\end{equation}
since the self-energy is determined from the two-particle vertex via Eqs.~\eqref{eq:KK-relations}. It means that  vertex  $\Lambda^{RA}_{\veck\veck'}(E;\omega,\mathbf{q})$  is directly related to the measurable macroscopic quantities only for $\omega=0$, $\vecq =\mathbf{0}$.    

The physical irreducible vertex $L^{ab}(E;\omega,\mathbf{q})$ can in this way be constructed to any approximate irreducible $\Lambda^{ab}(E;\omega,\mathbf{q})$ from the diagrammatic perturbation theory. Continuity equation is then saved and the density response displays a diffusion pole with a diffusion constant. The  isotropic \index{diffusion constant} diffusion constant is then expressed via a Kubo-like formula with the full two-particle vertex  \cite{Janis:2016aa}
\begin{multline}\label{eq:DC-static}
\pi n_{F}D =  \frac 1{N^{2}}\sum_{\veck,\veck'} \left\{( \hat\vecq\cdot\mathbf{v}_\veck)|G^{R}_\veck|^{2}\left[N\delta_{\veck,\veck'} + \Gamma^{RA}_{\veck\veck'}\lvert G^{R}_{\veck'}\rvert^{2}\right]
\right. \\ \left.
\times \left[\Im G^{R}_{\veck'}\hat\vecq\cdot\mathbf{v}_{\veck'}  + \Im\left(G^{R}_{\veck'}\hat\vecq\cdot\nabla_{\veck'}\Sigma^{R}_{\veck'}
  \right)\right] \Im \Sigma^{R}_{\veck'}\right\}\,,
\end{multline}   
where $\hat{\vecq}$ is the unit vector pointing in  the direction of the drifting electric force.
All the frequency variables are set zero, at the Fermi energy. This exact expression is the starting point for the derivation of consistent approximations for the diffusion constant needed  to reach quantitative results.

\section{Conclusions}

The coherent potential approximation was introduced and developed in the late sixties of the last century. Primarily it was restricted only to a quantum mechanical problem of a particle in a random lattice. The concept of a homogeneous coherent potential simulating the effect of the fluctuating environment so that locally and in the long-time limit there is no difference between the averaged and fluctuating environment proved useful beyond the one-particle quantum mechanics.      

The first step in giving the coherent potential a more general meaning was understanding the approximation in terms of Feynman diagrams. Using the functional integral as a generator of Feynman diagrams then allowed for transferring CPA onto many-body systems
with arbitrary local interaction or site-independent disorder. It appeared that CPA is another form of the cavity field in which all single-loop corrections to tree diagrams are summed. The final framing of CPA was achieved by introducing DMFT as the exact solution in the limit of quantum itinerant models in infinite lattice dimensions. 

The coherent potential approximation and its consistency and correct analytic behavior emerged within DMFT in a new light as an exact solution in a special limit. The equations determining the coherent potential deliver also a mean-field solution of interacting models with elastic scatterings, that is where the energy is conserved during the scattering events. Moreover, the thermodynamic formulation of the CPA equations for the disordered Anderson and Falicov-Kimball models  unveiled differences in the solutions of the two models for their response functions. It also gave a proper understanding of the Hubbrad-III approximation and rectified its inconsistency. 

The coherent potential approximation, unlike DMFT for the Hubbard model, is analytically solvable. It is a huge advantage, since it allows for the full analytic control of the mean-field behavior, including quantum criticality. Last but not least, CPA as a simpler DMFT may also serve as a test arena for reliability of approximations devised for the Hubbard model where it is otherwise uncontrolled. The major restriction in applicability of CPA is that the equilibrium states are not Fermi liquid and their analytic properties are not directly transferable to heavy-fermion systems.

Local mean-field approximations in whatever formulation are consistent only for the local variables or one-particle properties. Once we need to calculate the response of the extended system to an external perturbation, we must go beyond DMFT.  The non-local two-particle functions are not uniquely defined in DMFT, since the limit to infinite dimensions is not interchangeable with functional derivatives \cite{Janis:1999ab}. Then either the Ward identity or electron-hole symmetry is broken for non-local response functions.

Ambiguity in the mean-field definition of non-local two-particle Green functions reflects a severe problem of expansions around DMFT. In particular in case of CPA, the expansion around it fails to reproduce the Ward identity that is needed for the existence of the diffusion pole in the density response function and validity of macroscopic conservation laws. We proposed a solution to this problem in disordered systems which opens a new and more consistent framework to study systematically vanishing of diffusion and Anderson localization. A full solution of this problem for strongly correlated Fermi liquids is to be found.  

\section*{Acknowledgment}
I thank Jind\v rich Koloren\v c for his long-term fruitful collaboration on most of the results presented in this review as well as for his  help in preparation of the diagrammatic representation of the perturbation expansion in disordered systems. I acknowledge support from Grant No. 15-14259S of the Czech Science Foundation.   


\clearpage

\clearchapter


\end{document}